\documentclass[lettersize,journal]{IEEEtran}
\usepackage{booktabs}  
\usepackage{subcaption} 
\usepackage{multirow}
\usepackage{listings}
\usepackage{microtype} 
\usepackage{xargs}              
\usepackage{balance}
\usepackage{algpseudocode}
\usepackage{pifont}
\usepackage{mathtools}
\usepackage{amsmath}
\usepackage{url}
\usepackage{subcaption}
\usepackage{tabulary}
\usepackage{etoolbox}
\usepackage{mathrsfs}  
\usepackage[T1]{fontenc}
\usepackage{semantic}
\usepackage{url}
\usepackage{enumitem}
\usepackage{balance}
\usepackage{wrapfig}
\usepackage{bm}
\usepackage{pgf,tikz}
\usepackage[utf8]{inputenc}
\usepackage{fmtcount}
\usepackage{threeparttable}
\usepackage{pifont}
\usepackage{tcolorbox}
\usepackage{hyperref}
\usepackage{cleveref}
\usepackage{ulem}
\usepackage{amsthm}
\usepackage{tikz}
\usepackage{listings}
\usepackage{rotating}
\usepackage{amssymb}
\usepackage{orcidlink}
\usepackage{subcaption} 

\newcommand{\remark}[1]{
	\begin{tcolorbox}[
		boxsep=-0.5pt,
		standard jigsaw,
		boxrule=0.6pt,
		opacityback=0,
		sharp corners]
		#1
	\end{tcolorbox}
}

\definecolor{codeblack}{rgb}{0,0,0}
\definecolor{codegreen}{rgb}{0,0.6,0}
\definecolor{codegray}{rgb}{0.5,0.5,0.5}
\definecolor{codepurple}{rgb}{0.58,0,0.82}
\definecolor{backcolour}{rgb}{1,1,1}

\lstdefinestyle{mystyle1}{
	backgroundcolor=\color{backcolour},
	commentstyle=\color{mygreen},
	language={C}, 
	basicstyle=\ttfamily\fontsize{8}{9}\selectfont,
	keywordstyle=\fontsize{8}{9}\selectfont\bfseries\color{black},
	keywordstyle = [2]{\fontsize{8}{9}\selectfont\bfseries\color{blue}},
	otherkeywords = {},
	numbers=left,
	numberstyle=\tiny\color{codegray},
	stepnumber=1,
	numbersep=5pt,
	breakatwhitespace=false,           
	captionpos=b,                
	showspaces=false,                
	showstringspaces=false,
	showtabs=false, 
	tabsize=2,
	xleftmargin=2em
}

\newcommand{\mytodored}[1]{\textcolor{red}{\ding{46}~{\sf}~#1}}

\newcommand{\mytodocyan}[1]{\textcolor{cyan}{\ding{46}~{\sf}~#1}}
\newcommand{\mytodopink}[1]{\textcolor{purple}{\ding{46}~{\sf}~#1}}

\newif\ifshowcomments
\showcommentsfalse

\ifshowcomments
\newcommand{\wei}[1]{\mytodocyan{[Wei: #1]}}
\newcommand{\qhj}[1]{\mytodored{[qhj: #1]}}
\newcommand{\zhang}[1]{\mytodopink{[zhang: #1]}}
\else
\newcommand{\wei}[1]{}
\newcommand{\qhj}[1]{}
\newcommand{\zhang}[1]{}
\fi

\theoremstyle{definition}
\newtheorem{definition}{Definition}

\newtheorem{finding}{Finding}

\begin{document}



\title{Unveiling the Power of Intermediate Representations for Static Analysis: A Survey}

\author{Bowen Zhang\orcidlink{0000-0002-2985-6713}, Wei Chen\orcidlink{0000-0003-0335-5002}, Hung-Chun Chiu\orcidlink{0000-0002-1416-1409}, and Charles Zhang\orcidlink{0009-0006-8611-9468},\thanks{Bowen Zhang, Wei Chen, Hung-Chun Chiu, and Charles Zhang are with the Department of Computer Science and Engineering, The Hong Kong University of Science and Technology, Hong Kong, China (e-mail: bzhangbr@cse.ust.hk; wchenbt@cse.ust.hk;hchiuab@cse.ust.hk;charlesz@cse.ust.hk).}}

\markboth{Transactions on Software Engineering} {\textit{Zhang et al.}}


\maketitle
\begin{abstract}
Static analysis techniques enhance the security, performance, and reliability of programs by analyzing and portraiting program behaviors without the need for actual execution.
In essence, 
static analysis takes the Intermediate Representation (IR) of a target program as input to retrieve essential program information and understand the program.
However, 
there is a lack of systematic analysis on the benefit of IR for static analysis, 
besides serving as an information provider.
In general,
a modern static analysis framework should possess the ability to conduct diverse analyses on different languages,
producing reliable results with minimal time consumption,
and offering extensive customization options.
In this survey,
we systematically characterize these goals and review the potential solutions from the perspective of IR.
It can serve as a manual for learners and practitioners in the static analysis field to better understand IR design.
Meanwhile, numerous research opportunities are revealed for researchers.

\end{abstract}


\section{Introduction}
\IEEEPARstart{W}{ithout} actual program execution,
static analysis infers program properties to optimize program performance~\cite{olivo2015static,zhang2021sanrazor}, 
detect program defects and vulnerabilities~\cite{machiry2017dr,Lin2023ASAP,Fan2019Smoke},
assist program understanding~\cite{Sui2016SVF,Feng2015EXPLORER},
guide dynamic analysis~\cite{jeong2019razzer,li2017steelix}, etc.
Central to this technique is the utilization of Intermediate Representation (IR),
which provides the necessary information for the program.
Typically, 
to analyze a program, 
a static analyzer first acquires the corresponding IR as input. 
Subsequently, 
different analysis algorithms can be implemented by manipulating the IR,
making IR akin to a programming language for static analysis.

Throughout the years, 
the discipline of static analysis has experienced substantial advancements,
resulting in the creation of several theories and application scenarios.
Nevertheless, 
the specific manner in which IR can enhance a modern static analysis framework remains ambiguous.
They embody three overarching objectives of a static analysis framework.
Subsequently, we will provide a detailed demonstration of each of these objectives.

\textbf{Goal 1: Perform various analyses on different languages (Versatility).} 
A modern static analysis framework depends on the collaboration of multiple analyses,
for example,
memory leak detectors~\cite{Xie2005FSE, Fan2019Smoke} benefit from the assistance of call graph analysis and points-to analysis.
In addition, 
the prosperity in varied computational domains, such as machine learning, has motivated ambitious static analysis frameworks to attempt to repurpose their capabilities in these domains.
Meanwhile, these areas have also posed distinct analysis demands.
Hence, a framework must possess greater versatility in order to effectively handle the proliferation of analysis demands.

\textbf{Goal 2: Generate reliable results within a brief timeframe (Performance).}
The main emphasis of static analysis customers is on the quality of the analysis results.
If the memory leak detector discussed earlier generates a high number of false warnings or fails to detect a substantial amount of real bugs, users will grow frustrated and reluctant to study the bug reports.
In addition, users may also grow impatient if the framework does not effectively handle the time required to produce high-quality outcomes.
Hence, a framework must strike a balance between these two aspects to attain optimal performance.

\textbf{Goal 3: Offer friendly interfaces for customizing analysis tasks (Productivity).}
A framework should facilitate developers to perform effective secondary development.
For example, in the context of a previously described memory leak detection, 
it would have multiple practical specifications, 
such as labeling various versions of \textsf{malloc()} and \textsf{free()}. 
If the cost of customization is excessively high, 
the framework's popularity would decrease,
despite its provision of a theoretical foundation for such analysis.
Hence, a framework must be highly extensible and configurable to ease the burden of continuous development. 

Regrettably, 
within the present research community, 
it is still unclear whether IR can help achieve the three goals,
let alone the specific methodologies.
Specifically,
existing surveys and studies about IR primarily aim to serve various objectives related to compilation,
such as conducting optimization~\cite{Stanier2013IRImpertive, Chow2013IRSignificance} and representing parallel computing~\cite{Susungi2021IRParallel}.
As a result,
when encountering an issue related to IR in a static analysis framework,
researchers and practitioners can only refer to the compiler textbooks~\cite{Aho1986CompilersPT, EngineeringAComp},
which generally does not effectively help.

Worse,
some static analysis problems that could have been addressed through the lens of IR,
are not realized as well.
Lack of sufficient information about this question can lead to misconceptions. 
For instance, some people may confuse IR with a specific well-known instance such as LLVM IR. 
There are also viewpoints that consider IR merely as an input for static analysis, without any other critical roles.
Therefore,
this survey establishes the first ``manual'' for static analysis practitioners to wisely make decisions about IR in the context of static analysis.
Through reading this survey,
static analysis learners can also gain a better understanding of IR.
Furthermore,
numerous critical research opportunities are pointed out for future researchers to explore.

With the question \textbf{\textit{``How can IR benefit static analysis to achieve the goals?''}} in mind, we systematically examined the past works in the field of static analysis.
The involved publications encompass major conferences and journals in the areas of software engineering, programming languages, compilers, and security.
To provide a comprehensive view, 
we additionally surveyed some influential industrial static analysis frameworks in terms of their IRs.

The rest of this survey is organized as follows.
First,
we present several basic concepts and elaborate our study categories (\S~\ref{sec:prelim}).
Then,
we examine the existing works in each of the proposed categories (\S~\ref{sec:syntax}, \ref{sec:voc}, \ref{sec:query}, \ref{sec:preprocess}).
Finally,
we conclude the survey by delivering well-examined experiences to practitioners and leaving future researchers with research opportunities (\S~\ref{sec:discussion}).

\section{Preliminary}\label{sec:prelim}
In this section,
we provide the readers with fundamental background knowledge regarding the IR used in static analysis.
Specifically,
we begin by emphasizing the significance of IR. 
Subsequently, 
we elaborate our categories through an IR journey, covering some necessary concepts.

\subsection{Significance.}
The Intermediate representation (IR), 
serving as a program representation between the source code and the binary code, 
is necessary to achieve a modern static analyzer.
On the one hand, 
although the source code text of a program contains all the necessary program information for static analysis, 
the presence of complex and numerous language structures limits its capability to perform only text search-based static analysis.
For example, tools like \texttt{grep} struggle with tasks such as ``\textit{replacing the name of the class Foo with Bar}'',
because the various forms of class references all need to be considered, 
such as inheritance, variable definitions, and initialization function calls.
On the other hand, 
the binary representation of a program, 
such as \texttt{ELF} or \texttt{EXE} format,
 resides at a low level. 
They are specifically designed for hardware execution lacking high-level semantic information, 
which makes it impossible to achieve many static analyses.
In situations where only binary code is available, binary analysis becomes necessary, but it is beyond the scope of this survey.
Therefore, an IR is essential because 
it provides a structured representation of program semantics that is 
higher level than binary code 
and more parseable than source code text, 
making it ideal for static analysis.

\subsection{A Journey about IR.}

In this subsection,
we introduce our proposed categories by going through a journey of handling the IR in a static analysis framework.
The entire workflow is divided into the design and implementation stages.
\smallskip

\begin{figure}[t]
	\centering
	\begin{small}
	\begin{minipage}[t]{0.48\linewidth}
	\begin{align*}
		\textsf{Program} \ P \ &:= \  F^{+}\ \ v^{+} \\
		\textsf{Function} \ F \ &:= \ f\ (\ v^{+}\ ) \ \{ \ S^{+} \ \} \\ 
		\textsf{Statement} \ S \ &:= \ S_{decl} \ \ | \ \ S_{assign} \ \  | \ \ S_{if} \ \ | \ \ \cdots\\
		\textsf{Expression} \ E \ &:= \ E_{bin} \ \  | \ \  E_{call} \ \  | \ \ v \ \ | \ \ c \ \ | \ \ \cdots\\
		\textsf{Variable} \ v \ \ &\in \ V \ \ \ \ \  \textsf{Constant} \ c \ \ \in \ C \\
	\end{align*}
	\end{minipage}
	\hfill
	\begin{minipage}[t]{0.48\linewidth}
	\begin{align*}
	\textsf{Declaration} \ S_{decl} \ &:= \  v\ ::\ t\\
	\textsf{Assignment} \ S_{assign} \ &:= \  v \ \gets \  E \\
	\textsf{If Statement} \ S_{if} \ &:= \  \texttt{if} \ ( \ E\ )\  \{ S^{+}\}  \\
	\textsf{Binary Operation} \ E_{bin} \ &:= \  E + E \ | \ E - E \ | \cdots  \\
	\textsf{Function Call} \ E_{call} \ &:= \  F \ ( \ E^{+} \ )  \\
	\textsf{Type} \ t \ \ &\in \ T  \\
	\end{align*}
	\end{minipage}
    \end{small}
	\vspace*{-3mm}
	\caption{Program Concepts}
	\label{fig:program-concepts}
\end{figure}

\textbf{Stage 1: Design.} 
The essence of IR design is the design of a language.
Consider a typical formulation of the program in Fig.~\ref{fig:program-concepts},
we can understand the involved concepts from two perspectives:
\begin{itemize}
\item
Vertically, 
the program's functionality is primarily encapsulated within \textit{functions}. 
The \textit{statements} within these functions represent the program's behavior,
while the \textit{expressions} within the statements denote composite operations. 
Furthermore, 
the atomic concepts such as 
\textit{variables}, \textit{constants}, and \textit{types} collectively form the program's memory model.
\item
Horizontally, 
these concepts can be instantiated into various variants, 
such as statements encompassing declarations, assignments, if statements, and more. 
Such diversity of variants contributes to the program's richness in terms of features.
\end{itemize}

These two perspectives essentially correspond to the syntax and vocabulary of a program. 
Therefore,
the design of an IR aims to reshape the \textbf{IR syntax} and \textbf{IR vocabulary} in a way that is tailored for static analysis purposes for the target programs.
As a result,
this stage defines an IR in terms of data structure.
Meanwhile,
an IR library is developed in accompany to support various basic functionalities such as reading and serializing IR, 
as well as manipulating IR elements.
\smallskip

\textbf{Stage 2: Implementation.}
This stage turns the designed IR into practical and useful.
Currently, 
the IR we have designed exists in a vacuum.
On the one hand,
even if we can utilize the provided IR library, the logic of static analysis remains highly complex.
Therefore,
many studies utilize the raw IR library to further encapsulate higher-level interfaces that make it easier to describe static analysis algorithms.
Such interfaces are referred to as \textbf{IR query}.
On the other hand,
we need to obtain a ``good'' IR program to feed into the static analyzer.
Hence,
researchers have proposed different kinds of \textbf{IR preprocessing} techniques, 
including unifying the capture of IR programs from different languages, simplifying the IR, and reducing the size of the IR.
As a result,
this stage produces two kinds of middleware libraries,
one for preprocessing the IR and another for providing IR queries to ease the burden of implementing static analysis.
\smallskip

In the subsequent article, 
we first introduce IR design in terms of IR syntax (\S~\ref{sec:syntax}) and IR vocabulary (\S~\ref{sec:voc}),
then we illustrate IR implementation via IR query (\S~\ref{sec:query}) and IR preprocessing (\S~\ref{sec:preprocess}).

\section{IR Syntax}\label{sec:syntax}
IR syntax refers to the organization of program elements.
The central problem in designing it is to determine which program relations to select for emphasis.
Over the past few decades,
both academia and industry have extensively investigated this topic and put forth numerous IR syntaxes.
In what follows,
we will present these syntaxes based on four primary categories of program relationships (\S~\ref{subsec:syntax}, \ref{subsec:control}, \ref{subsec:eq}, \ref{subsec:dep}).
By comprehending the underlying principles and benefits of these syntaxes,
readers can make well-informed choices when choosing among existing syntaxes or extending them to suit their specific analysis purposes.
Specifically,
for several typical syntaxes,
we will provide examples of them all based on the source code in Fig.~\ref{fig:fib},
which is an imperative implementation of Fibonacci.

\begin{figure}[t]
\centering
\lstset{style=mystyle1}
\begin{lstlisting}
int fib(int n) {
    int result = 0;
    if (n == 1) 
        result = 1;
    if (n > 1) 
        result = fib(n-1) + fib(n-2);
    return result;
}
\end{lstlisting}
\vspace*{-2mm}
\caption{A code example of calculating Fibonacci}
\label{fig:fib}
\end{figure}

\subsection{Syntactic Nesting Relation}
\label{subsec:syntax}
The first program relation to introduce is the syntactic nesting relation (Def.~\ref{def:syntax-nesting}).
It abstracts away the tedious details in the grammar of a program, such as parentheses and curly brackets,
helping us grasp the most essential information about a program element from a syntactical perspective.
By extracting syntactic nesting relations during the parsing process of source code~\cite{Aho1986CompilersPT},
and using them as edges connecting corresponding program elements,
the Abstract Syntax Tree can be formed,
as shown in Def.~\ref{def:ast}.

\begin{definition}[Syntactic Nesting Relation]\label{def:syntax-nesting}
The syntactic nesting relation is a binary relation among program elements like functions, statements, expressions and variables.
Two elements $e_1$ and $e_2$ exhibit this relation iff $e_2$ is directly nested within $e_2$,
i.e. in the grammar $e_1$ directly contains $e_2$ without other intermediate program elements.
\end{definition}

\begin{definition}[Abstract Syntax Tree]\label{def:ast}
The Abstract Syntax Tree (AST) is a tree structure $T = (V, E)$,
where:
\begin{itemize}
	\item The node set $V$ represents the program elements, such as function, statement, expression, etc.
	\item The edge set $E \subset V \times V$ represents the syntactic nesting relations between program elements. An edge $(v_1, v_2) \in E$ denotes that $v_2$ is directly nested within $v_1$ according to Def.~\ref{def:syntax-nesting}.
\end{itemize}
\end{definition}

\begin{figure*}[h]
	\centering
	\begin{subfigure}[b]{0.35\textwidth}
		\centering
		\includegraphics[width=\textwidth]{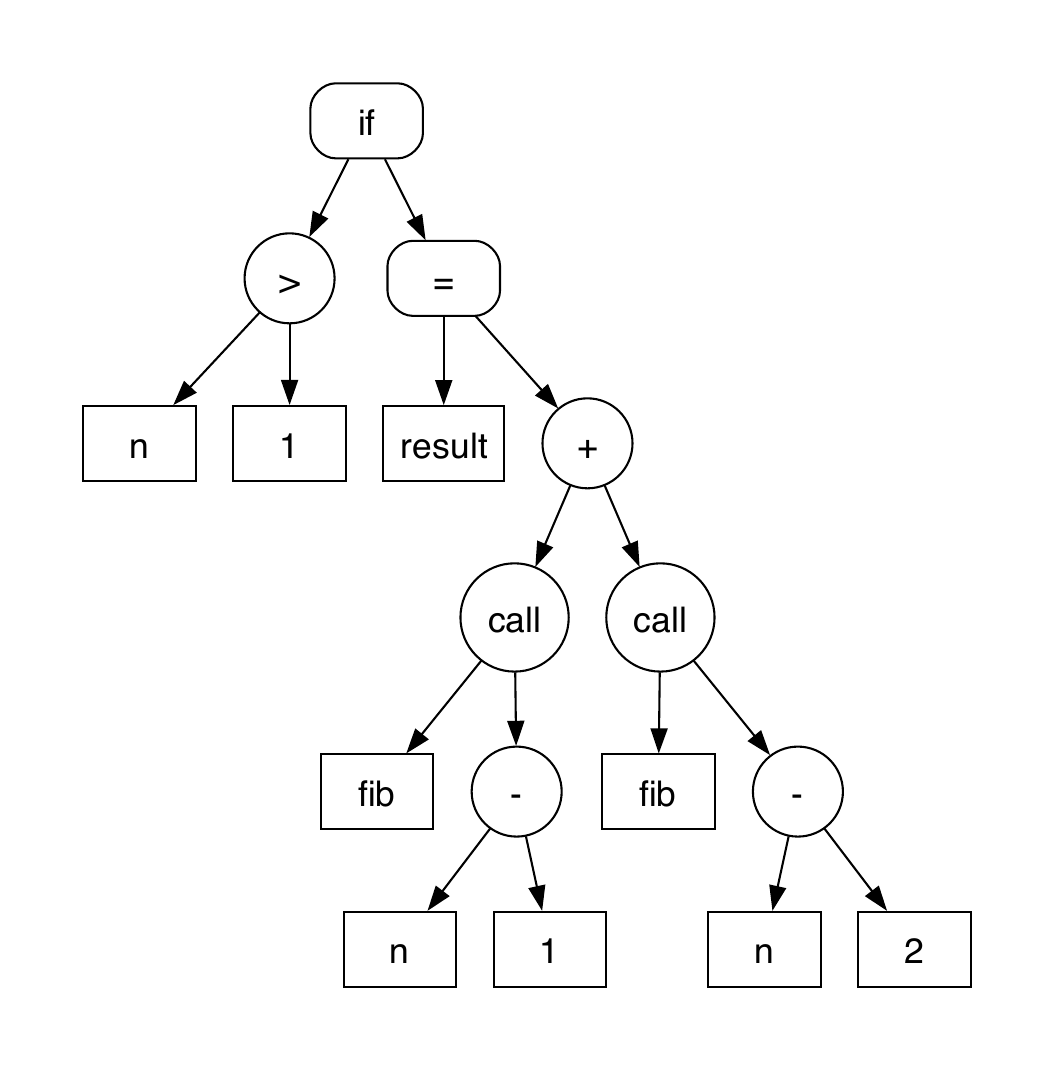}
		\label{fig:ast}
		\vspace*{-6mm}
		\caption{}
	\end{subfigure}\hspace*{-1cm} 
	\begin{subfigure}[b]{0.35\textwidth}
		\centering
		\includegraphics[width=\textwidth]{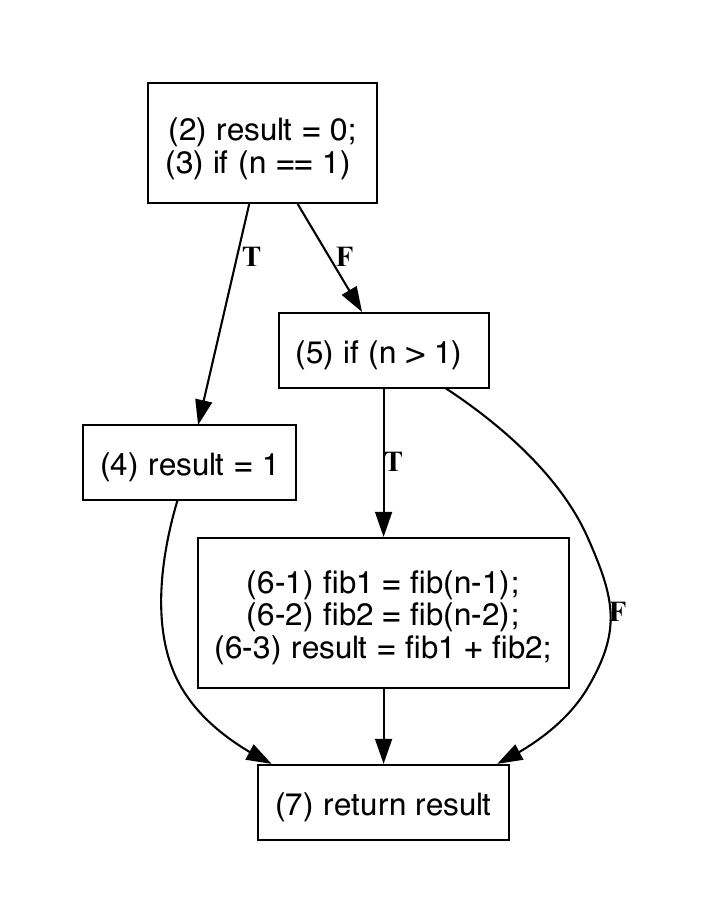}
		\label{fig:cfg}
		\vspace*{-6mm}
		\caption{}
	\end{subfigure}\hspace*{-1cm} 
	\begin{subfigure}[b]{0.4\textwidth}
		\centering
		\includegraphics[width=\textwidth]{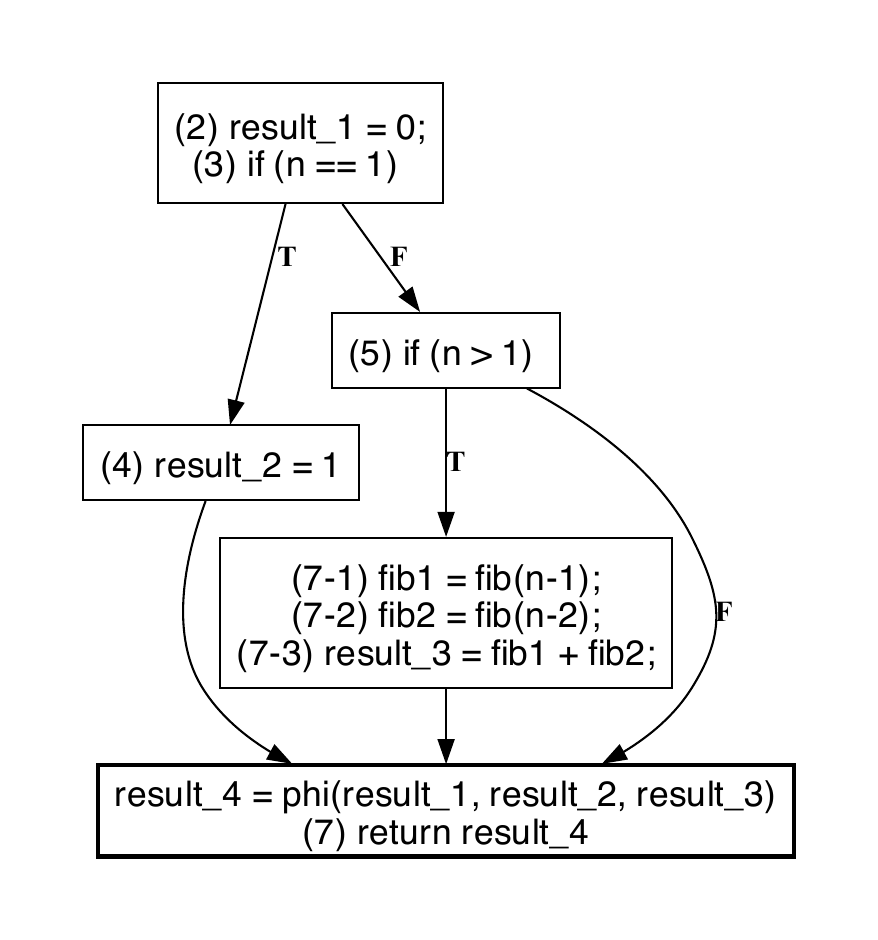}
		\label{fig:ssa}
		\vspace*{-6mm}
		\caption{}
	\end{subfigure}
	\caption{The AST, CFG, and SSA corresponding to the code in Fig.~\ref{fig:fib}}
	\label{fig:ast_cfg_ssa}
\end{figure*}

Fig.~\ref{fig:ast_cfg_ssa}(a) demonstrates the AST corresponding to lines 5-6 in Fig.~\ref{fig:fib}.
Specifically,
the node shapes differentiate between statements (e.g. \textsf{if}), expressions(e.g. \textsf{>}), and identifiers(e.g. \textsf{a}),
while the edges highlight the scopes within the source code.
Such hierarchical IR syntax, 
which closely mirrors the source code,
is well suited for performing tree traversal style analysis on the nested scopes,
such as performing type analysis~\cite{Avik2017oopsla} or matching patterns for try-catch blocks.

\subsection{Control Flow Order}
\label{subsec:control}
The second program relation is control flow order,
which is defined as follows:
\begin{definition}[Control Flow Order]\label{def:flow-order}
The control flow order of a program is a binary relation concerning program statements in the same function.
Two statements $s_1, s_2$ satisfy this relation iff during the program execution,
there is a potential path from $s_1$ to $s_2$.
Further,
if $s_2$ can be executed right following $s_1$,
we say there is a direct control flow order between them.
\end{definition}

Overall,
the control flow order provides an opportunity for us to reason program facts related to program execution,
such as typestate properties~\cite{typestate}.
For the classic abstract interpretation algorithm~\cite{Cousot1977AI}, 
control flow order is also the foundation of its implementation.
Recall that in the design of the AST, 
the best effort we can provide for control flow order is to maintain the order among statements located at the same level in the tree, 
based on their ordering in the source code writing level.
Obviously,
such an order is not explicit enough.
Next,
we start to develop new IR syntaxes to better represent control flow order.

To begin with,
we can traverse the AST to identify control flow statements, such as \textsf{if}, \textsf{while}, and \textsf{break},
and determine the potential statements they may jump to,
thus forming a set of edges.
These edges, combined with the statement order in the source code, collectively form the statement-level control flow graph (CFG),
as described in Def.~\ref{def:cfg1}.

\begin{definition}[Statement-level Control Flow Graph]\label{def:cfg1}
Given a function $f$ and its AST $T = (V,E)$,
the corresponding Control Flow Graph (CFG) is a graph $G = (S, E)$, where:
\begin{itemize}
\item The node set $S \subset V$ denotes the set of program statements in $f$.
\item The edge set $E \subset S \times S$ denotes the control flow edges at the statement level. 
An edge $(s_1, s_2) \in E$ represents a possible direct control flow order between $s_1$ and $s_2$.
\end{itemize}
\end{definition}

However, the control flow order expressed by the current CFG remains obscure.
Firstly,
the execution order between expressions is not reflected.
More importantly,
the combination of nested statements and the diverse behavior of control flow statements make it challenging to determine clear ``boundaries'' between sequentially executed portions in the graph.
This increases the difficulty of maintaining analysis results in flow-sensitive analysis.
At the same time,
preserving the syntactic scoping relation hinders the simplification of control flow,
such as eliminating unreachable code.

To overcome these limitations,
a modern control flow analysis~\cite{Allen1970CFA} would flatten the nested scopes, 
consolidate the statements and expressions into \textit{instructions} (operations),
and group them into \textit{basic blocks}.
Therefore,
we obtain an enhanced CFG in Def.~\ref{def:cfg2}, 
which is not associated with the original AST.

\begin{definition}[Control Flow Graph]\label{def:cfg2}
Given a function $f$,
its Control Flow Graph (CFG) is a graph $G = (I, B, E)$, 
where:
\begin{itemize}
\item The set $I$ represents instructions, which are the atomic operations within $f$.
\item The node set $B$ represents basic blocks.
	Each block $b \in B$ is a list of instructions $(i_1, i_2, \dots, i_n)$ where $i \in I$ that are always executed sequentially.
\item The edge set $E \subset B \times B$ denotes the control flow edges between basic blocks. 
	An edge $(b_1, b_2) \in E$ denotes there is a control flow order relation between the last instruction of $b_1$ and the first instruction of $b_2$.
\end{itemize}
\end{definition}

The instruction,
also known as the three-address code~\cite{Aho1986CompilersPT},
serves as the basic semantic unit of the IR.
An instruction typically performs an atomic action,
such as arithmetic or memory operations.
It can be stack-based~\cite{Gosling1995Bytecode} or register-based~\cite{Davidson1980Peephole}.
A basic block is a single-entry, single-exit structure,
containing several instructions that must execute sequentially. 
Fig.~\ref{fig:ast_cfg_ssa}(b) shows an example of register-based CFG corresponding to Fig.~\ref{fig:fib}.
Specifically, we can identify three possible execution paths in the function by enumerating the paths from the entry block (top) to the exit block (bottom) in the graph.

By using CFG as IR syntax,
the flow-sensitive analysis can be greatly facilitated.
For instance,
a typical forward dataflow analysis maintains the program state at the start and end of each block.
Within a block,
the analysis engine obtains the state at the start of the block,
applies a flow function based on each instruction, 
and updates the program state at the end of the block for propagation to subsequent blocks.
This convenience stems from the organizational approach of explicitly representing control flow order through blocks and instructions.

Moreover,
in the presence of function calls, 
the control flow order exhibits a specialized relationship known as the call relation.
It can be viewed as the control flow order between a function call $c$ and the entry point of the called function $f$.
The CFG also needs to represent this to support inter-procedural analysis.
For direct function calls, 
obtaining such relationships requires no effort.
However, the presence of language features like function pointers~\cite{cg-fpointer}, lambda functions~\cite{cg-functional}, inheritance, and overloading~\cite{cg-oop} introduces indirect function calls, 
making the acquisition of such relationships non-trivial.
After the resolution of call relation,
the IR syntax with enhanced call-edges is called Inter-procedural Control Flow Graph (ICFG),
as defined as follows:

\begin{definition}[Inter-procedural Control Flow Graph]\label{def:icfg}
Given a program $p$ with a function set $F$,
its Inter-procedural Control Flow Graph (ICFG) is a graph $G^{*} = (\mathcal{G}, C)$,
where:
\begin{itemize}
\item $\mathcal{G}_f = (I_f, B_f, E_f)$ represents the intra-procedural CFG defined in Def.~\ref{def:cfg2}, for each function $f \in F$.
\item $C \subset \{I_f | f \in F\} \times F$ is the set of call-edges recording the call relation. An edge $(i, f) \in C$ denote a call instruction and its potential callee function $f$.
\end{itemize}
\end{definition}

Based on ICFG, 
if we represent functions as nodes and the call relation among functions as edges, 
we obtain a commonly used notion called the Call Graph (CG).
To perform inter-procedural analysis,
there are two mainstream approaches.
The bottom-up approach~\cite{Barth1978Dataflow} computes summaries for callees and applies them when handling the callers,
following the inverse topological order provided by the CG.
The top-down approach~\cite{Cousot1977AI, lu2021k-cfa} simply ``inlines'' the callee function when reaching the call site.

\subsection{Equivalence}\label{subsec:eq}
In this section, 
we unveil how to leverage varying degrees of equivalence relations in the IR to optimize the syntaxes discussed earlier, 
thereby obtaining several impactful IR syntaxes.
The overall idea is to ``memorize'' the equivalence relations within the IR, 
thus relieving subsequent static analysis from having to maintain this information.

\subsubsection{Syntactic Equivalence}

For the AST,
the optimization opportunity lies in recognizing syntactic-level equivalence, 
as depicted below:

\begin{definition}[Syntactic Equivalence]\label{def:syntax-eq}
Two program elements of the same syntactic kind are considered syntactically equivalent iff their structure and constituents are recursively equivalent.
\end{definition}

Essentially, 
syntactic equivalence manifests as isomorphic subtrees in the AST.
By merging these syntactically equivalent subtrees,
the notion of Direct Acyclic Graph (DAG~\cite{Davidson1980Peephole}) is introduced.
In particular,
the construction of DAG involves memoizing the syntactic elements created during AST construction and reusing them when they reappear.
In compilers, the benefits of using a DAG typically include compact storage and better support optimizations such as constant propagation. 
As for static analysis, a DAG can assist in identifying the usage of the same variable/expression at different locations without the need for matching on the AST.
Moreover,
DAG facilitates analyses based on syntactic similarities, such as Type I and II code clones~\cite{VUDDY, CP-Miner}.

\subsubsection{Value Equivalence}
\label{subsubsec:value-eq}

Further,
we define the value equivalence as follows:
\begin{definition}[Value Equivalence]\label{def:value-eq}
Two program elements $o_1$ and $o_2$ are considered value equivalent iff they can produce the same value in all executions.
\end{definition}

Many compiler optimizations such as common subexpression elimination and copy propagation are used to identify value equivalence.
Exposing value equivalence in the IR will be beneficial to down-stream static analysis tasks,
as it further eliminates ambiguity between program elements at different locations in the program.
To achieve this goal,
researchers have established the Static Single Assignment form (SSA~\cite{Cytron1989SSA, Cytron1991SSA}),
which can be viewed as CFG enhanced with value equivalence.
SSA has a simply definition as below:
\begin{definition}[Static Single Assignment Form]\label{def:ssa}
A program conforms to static single assignment form iff each variable in it is assigned only once, i.e. at its definition point.
\end{definition}

This means that variables in SSA are not reassigned,
which achieves the aforementioned goal of eliminating ambiguity.
Moreover,
SSA explicitly represents the def-use chain,
as the uses of each variable can transparently locate their definition points.
This significantly facilitates sparse dataflow analysis~\cite{Oh2012sparse, Hardekopf2011sparse},
which propagates dataflow facts ``sparsely'' through the def-use chain, instead of control flow edges.

The construction of SSA typically requires performing forward dataflow analysis on the CFG.
First,
variable reassignments are identified and renamed with new and unique variable names,
which can be regarded as a lightweight global value numbering (GVN~\cite{Aho1986CompilersPT}).
Second,
at control flow merging points,
some special $\phi$ functions are inserted to recognize the values of the same variable from different control flows.
Specifically,
a $\phi$ function $x_{\phi} \gets \phi(x_{b_1},x_{b_2}, \dots)$ implies that executions from different predecessor blocks $b_1, b_2, \dots$ correspond to different values.
An illustrative example of SSA form is shown in Fig.~\ref{fig:ast_cfg_ssa}(c).
It can be observed that compared with the original CFG in Fig.~\ref{fig:ast_cfg_ssa}(b),
the different definitions of variable \textsf{result} are renamed (\textsf{result\_1}, \textsf{result\_2}, and \textsf{result\_3}).
Subsequently,
a $\phi$ function is placed at the exit block which receives the three different values. 

Furthermore,
the number of $\phi$ functions in SSA can be effectively reduced by only inserting them at the dominance frontier and performing live variable analysis.
This leads to the notion of pruned SSA~\cite{Choi1991PrunedSSA}.

The advent of SSA has led to numerous variants that accommodate different purposes of analysis.
Specifically, while SSA captures the def-use relationships of top-level variables, 
Hash SSA (HSSA~\cite{Chow1996HSSA}) considers the aliasing relationships of pointers.
Then, Single Static Information (SSI~\cite{Ananian2002SSI, Boissinot2012SSI}),
a ``symmetrical'' syntax of SSA, 
captures the use-def chain to support backward sparse dataflow analysis better.
Lastly,
Gated SSA (GSA~\cite{Havlak1993GSA}) extends the $\phi$ function to represent predicates, 
allowing for differentiation of values under different pre-conditions.
Overall, the core idea behind these variants is also to explicitly represent value equivalence on the CFG.

\subsection{Dependence}
\label{subsec:dep}
So far the IR syntaxes in \S~\ref{subsec:syntax} and \S~\ref{subsec:control} mainly organize the IR based on the relative positioning among program elements (nesting, ordering).
Moreover,
the optimizations in \S~\ref{subsec:eq} expose equivalence,
but are still under such a framework.
However,
as researchers delve deep into the behaviors of programs,
they have realized that the understanding of some part of code often depends on another piece of code.
Such dependency relations cannot be captured by the previous syntaxes,
and have prompted researchers to design various dependency graph-based IR syntaxes.

\subsubsection{Data Dependence}
The first dependency relation is data dependence, defined as follows:
\begin{definition} [Data Dependence]
\label{def:dd}
Given two program elements $e_1$ and $e_2$ where $e_1$ can execute before $e_2$,
$e_2$ is data dependent on $e_1$ iff $e_2$ takes the data produced by $e_1$ as an operand.
\end{definition}

Indeed,
data dependence and the def-use relation mentioned in \S~\ref{subsubsec:value-eq} essentially correspond to the same ``edge set'' for a program.
The only distinction is that def-use relation focuses on variables as its objects,
whereas data dependence focuses on program statements.
By employing the data dependence as edges to organize the program,
the Data Dependence Graph (DDG) is established.
with the following definition:

\begin{definition} [Data Dependence Graph]
\label{def:ddg}
The data dependence graph of a program is a graph $G = (V, E)$, where:
\begin{itemize}
\item The node set $V$ represents operators and operands.
\item The edge set $E \subset V \times V$ represents data dependence edges. An edge $(v_1, v_2) \in E$ denotes an operator $v_1$ that uses the data produced by $v_2$, which can either be an operand or an operator. 
	\end{itemize}
\end{definition}

\begin{figure}[t]
\begin{minipage}{0.5\textwidth}
\centering
\includegraphics[width=\textwidth]{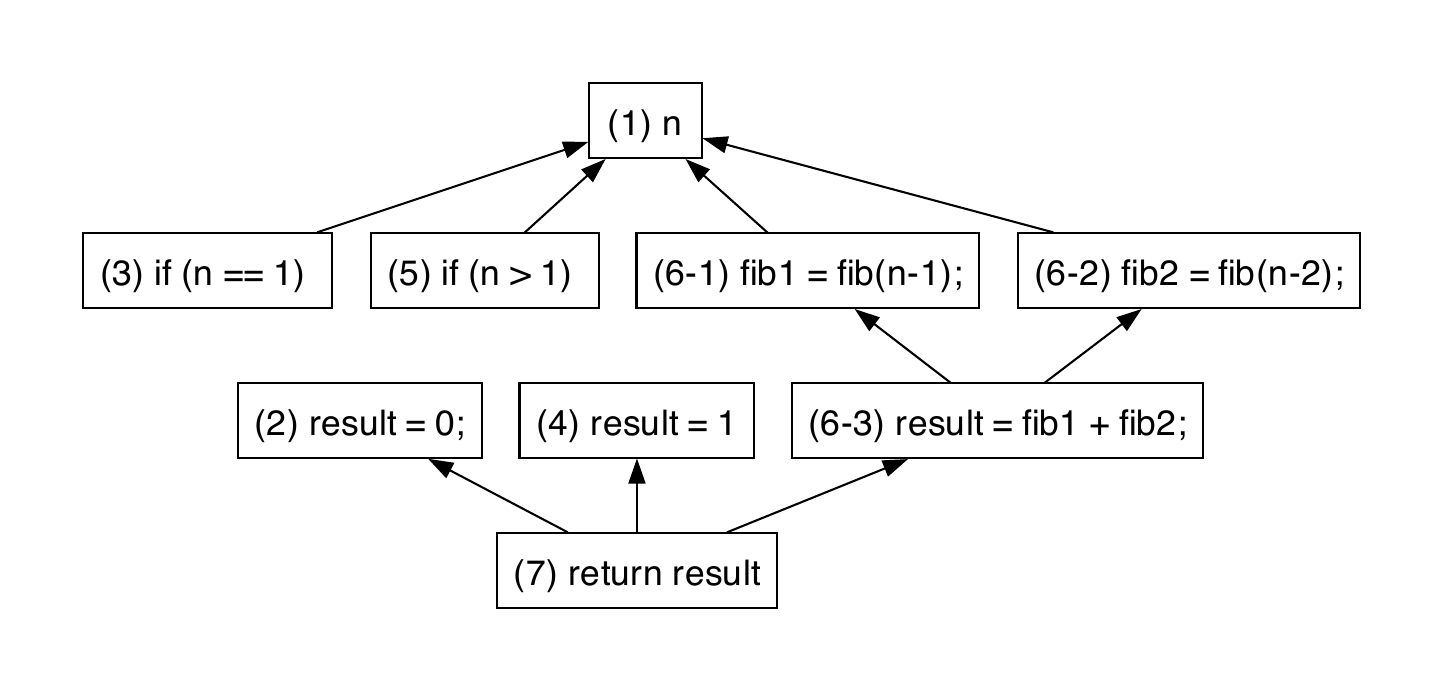}
\caption{The DDG for code in Fig.~\ref{fig:fib}}
\label{fig:ddg}
\end{minipage}
\end{figure}

The core spirit of DDG lies in the introduction of operands and operators,
which represent a program based on its behaviors on data production and usage.
Specifically, 
at the AST level,
constants typically serve as operands, 
expressions participate as both operands and operators, 
while statements generally only act as operators. 
On the CFG level, 
instructions are typically used as operators while also potentially serving as operands.
Based on an AST or a CFG,
the construction of DDG mainly involves performing data flow analysis to collect the data dependence edges between these operands and operators~\cite{Harrod1993PGDCons}.
Fig.~\ref{fig:ddg} shows the DDG for the Fibonacci program in Fig.~\ref{fig:fib}.
The nodes include the CFG instructions in Fig.~\ref{fig:ast_cfg_ssa}(b) as well as the parameter \textsf{n}.
Each directed edge represents an operator-operand pair.
For instance,
the edge connecting the node labeled \textsf{6-3} to the node labeled \textsf{6-2} indicates the usage of recursion result within the addition.
Specifically, constants such as 0, 1, and 2 are not treated as nodes in the example. 
This depends on the specific implementation of the DDG.

The organization of DDG allows us to group together program elements that are related through data usage,
by efficient graph traversal.
The \textit{slice-style} analysis can be easily implemented by traversal through the data dependence edges.
For example,
in Fig.~\ref{fig:ddg},
to find all operations involved in the returned result (label \textsf{7}),
we can use a BFS to sequentially identify the nodes $\{\textsf{2}, \textsf{4}, \textsf{6-3}, \textsf{6-1}, \textsf{6-2}, \textsf{1}\}$.
Moreover,
the \textit{taint-style} analysis can be facilitated by a propagation through the reversed data dependence edges.
For instance,
in Fig.~\ref{fig:ddg},
to find the operations involving the parameter \textsf{n} (label \textsf{1}),
we can use a BFS to sequentially identify the nodes  $\{\textsf{3}, \textsf{5}, \textsf{6-1}, \textsf{6-2}, \textsf{6-3}, \textsf{7}\}$.

The earliest prototype of DDG was named the Data Flow Graph~\cite{Dennis1980DFG},
which was used to describe a computing architecture.
Since then, researchers have been continuously expanding the family of DDGs. 
Essentially, this expansion represents a deeper exploration of the essence of data dependence.
Kuck et al~\cite{Kuck1981DDGCompiler} explore several dependency relations derived from data dependence,
which are described as follows:
\begin{definition}[Derived Data Dependences]
\label{def:derived-dep}
Given two program elements $e_1$ and  $e_2$ where $e_1$ can execute before $e_2$,
we define three dependency relations as follows:
\begin{itemize}
	\item Anti- Dependence: $e_2$ has anti- dependence on $e_1$ iff the variable written by $e_2$ has been read by $e_1$.
	\item Output Dependence: $e_2$ has output dependence on $e_1$ iff they write the same variable.
	\item Input Dependence: $e_2$ has input dependence on $e_1$ iff they read the same variable.
\end{itemize}
\end{definition}

Identifying anti- and output dependence is crucial for scheduling instruction pipeline execution in hardware,
and obviously, they do not apply to programs in SSA form.
Moreover,
input dependence connects multiple usages of the same variable,
offering a ``shortcut'' for analyzing the behavior of API usages~\cite{Zhang2014APIDG, Lin2023ASAP},
as a sequence of related API calls share the input parameter.

Certain language features involve data dependencies that are not directly manifested at the top level of the IR,
such I/O operations and pointer arithmetic.
Therefore,
some studies enhance the DDG by adding additional nodes and edges to explicitly represent such data dependencies. 
To model I/O operations,
the Value State Dependence Graph (VSDG~\cite{Johnson2003VSDG}) introduces nodes to represent memory regions and connects them to the nodes that store/load them,
through a new type of edge called \textit{state dependence edge}.
Similarly,
the Symbolic Expression Graph (SEG~\cite{Shi2018Pinpoint}) treats memory objects identified by pointer analysis as new nodes to connect pointer references and dereferences.
To further handle pointer side effects across functions,
some \textit{pseudo nodes} that present parameters and arguments are generated  and connected at call sites through mod-ref analysis.

Since DDG emphasizes data dependence without encoding control flow-related information, 
it is unable to fully represent the execution behavior of a program.
Hence,
another line of research aims to enhance the execution semantics within the framework of the DDG.
Specifically,
the Value Dependence Graph (VDG~\cite{Weise1994VDG}) employs a functional approach to represent control structures in programs.
For instance,
in VDG,
loops are converted to tail recursion, and the branches are converted to conditional operations.
Moreover,
the Dependence Flow Graph (DFG~\cite{Pingali1991DFG}) extends DDG to a more complicated syntax, Petri-net, 
to represent the program's state through the flow of tokens.

\begin{figure*}[t]
	\centering
	\begin{minipage}[b]{0.40\textwidth}
		\centering
		\includegraphics[width=0.8\textwidth]{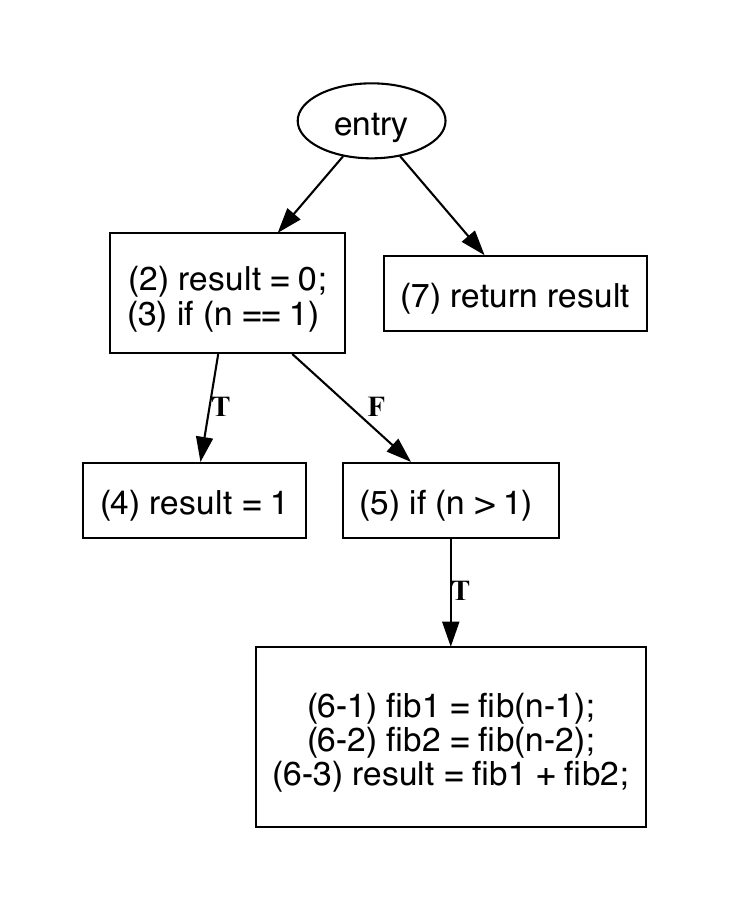}
		\caption{The CDG for code in Fig.~\ref{fig:fib}}
		\label{fig:cdg}
	\end{minipage}\hfill
	\begin{minipage}[b]{0.40\textwidth}
		\centering
		\includegraphics[width=\textwidth]{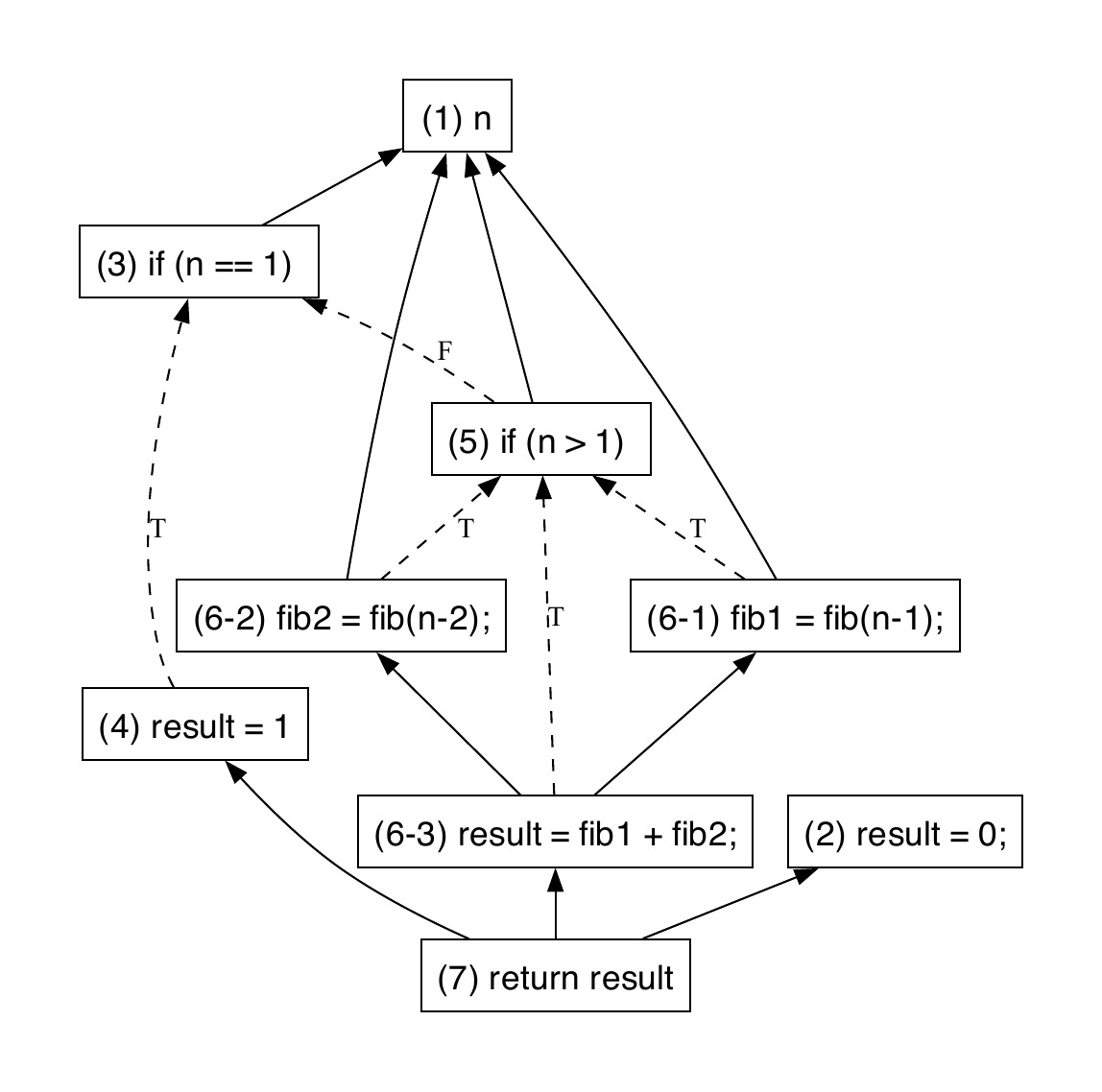}
		\caption{The PDG for code in Fig.~\ref{fig:fib}}
		\label{fig:pdg}
	\end{minipage}
\end{figure*}

\subsubsection{Control Dependence}
The other dependency relation is the control dependence,
which is defined as follows:

\begin{definition}[Control Dependence]
\label{def:cd}
Given two statements $s_1, s_2$,
$s_1$ has control dependence on $s_2$ if the execution of $s_1$ depends on the outcome of $s_2$.
\end{definition}

An alternative way to understand Def.~\ref{def:cd} is that:
\begin{itemize}
	\item there exists an execution path from $s_1$ through $s_2$ to the program exit.
	\item there also exists an execution $s_1$ to the program exit skipping $s_2$.
\end{itemize}
From this perspective,
it becomes clear that control dependence essentially records the the preconditions upon which the execution of a statement depends.
This makes it vital for understanding the program behaviors.

Based on control dependence,
researchers have established the Control Dependence Graph (CDG),
depicted in Def.~\ref{def:cdg}.
It is constructed through the computation of dominance frontiner~\cite{Cytron1991SSA, Harrod1993PGDCons} which determines the direct control dependence edges.

\begin{definition}[Control Dependence Graph]
\label{def:cdg}
The control dependence graph of a program is a graph $G = (V, E)$, where:
\begin{itemize}
	\item The node set $V$ includes statements or regions. Specifically, a region organizes the nodes that share the same control dependence.
	\item The edge set $E$ represents direct control dependencies between nodes.
\end{itemize}
\end{definition}

Fig.~\ref{fig:cdg} shows an example of CDG.
Specifically,
the \textsf{entry} region organizes all the nodes at the top level,
while the other nodes correspond to the basic blocks in Fig.~\ref{fig:ast_cfg_ssa}(b).
Each edge represents a direct control dependence,
and its edge label further indicates the condition.
By backtracking through these edges, we can gain a clear understanding of the essential preconditions of the program execution.
For example,
the return instruction with label \textsf{7} will execute under any condition,
as it is directly connected to the \textsf{entry}. 
In addition,
the instruction \textsf{result = 1} with label \textsf{4} will execute under the condition \textsf{n==1}.

However,
the CDG itself cannot fully support path-sensitive analysis.
Considering the obtained condition \textsf{n==1} in the last example,
it is still unknown whether there are further constraints on the variable \textsf{n}.
To overcome this barrier,
researchers have cleverly organized control dependence and data dependence into the same graph.
The resulting IR syntax is known as the Program Dependence Graph (PDG~\cite{Ferrante1987PDG, Horwitz1992PDGSE}).
The construction of PDG involves building subgraphs for both DDG and CDG.
Specifically,
nodes in DDG are linked to their corresponding nodes in the CDG through the their scopes or basic blocks in AST or CFG.
PDG can be understood as an on-demand structure for path condition~\cite{Shi2021Fusion},
as its compuation can be naturally performed through a recursive traversal procedure on PDG utilizing two kinds of edges.
As shown in Fig.~\ref{fig:pdg},
by utilizing the data dependence edge from the node with label \textsf{3} to the node with label \textsf{3},
we can ultimately understand that the variable \textsf{n} is a unconstrainted value.
Furthermore,
PDG has propelled the development of similarity-based static analysis~\cite{Gabel2008clone, sas01slice} by enabling them to comprehend deeper program semantics. 

As a higher-order syntax, 
the quality of the PDG could be influenced by the syntax from which it is derived.
The construction of the CDG subgraph would be more effective when based on CFG due to the explicit control flow information.
Additionally, 
the benefits brought by SSA form can also help improve the DDG.
This motivation led to the design of the sea-of-nodes (SoN~\cite{Click1995SoN}). 
Similarly, another approach called Program Dependence Web (PDW~\cite{Ottenstein1990PDW}) leverages GSA to better express paths.
Finally, the inter-procedural extension of PDG, the System Dependence Graph (SDG~\cite{Horwitz1988SDG}), 
will also benefit from more accurate call graph construction.

\subsection{Choice of IR Syntax}

\begin{table*}[h]	
	\centering
    \renewcommand\arraystretch{1.2}
    \begin{tabular}{|c|l|l|l|l|l|}
    \hline
    \multicolumn{1}{|c|}{} & \multicolumn{1}{c|}{\textbf{Syntax}} & \multicolumn{1}{c|}{\textbf{Order}} & \multicolumn{1}{c|}{\textbf{Equivalence}} & \multicolumn{2}{c|}{\textbf{Dependence}} \\ \cline{5-6}
    \multicolumn{1}{|c|}{} & \multicolumn{1}{c|}{} & \multicolumn{1}{c|}{} & \multicolumn{1}{c|}{} & \multicolumn{1}{c|}{\textbf{Data}} & \multicolumn{1}{c|}{\textbf{Control}} \\ 
    \hline
    \textbf{AST} & \checkmark &                     &                              &                      &                    \\ \hline
    DAG          & \checkmark &                     & \checkmark (Syntax)          &                      &                    \\ \hline
    CFG          &            & \checkmark          &                              &                      &                    \\ \hline
    ICFG         &            & \checkmark (inter-) &                              &                      &                    \\ \hline
    \textbf{SSA} &            & \checkmark          & \checkmark (Value)           &                      &                    \\ \hline
    HSSA         &            & \checkmark          & \checkmark (Value+pointer)   &                      &                    \\ \hline
    SSI          &            &                     & \checkmark (Value+backward)  &                      &                    \\ \hline
    GSA          &            &                     & \checkmark (Value+condition) &                      &                    \\ \hline
    \textbf{DDG} &            &                     &                              & \checkmark           &                    \\ \hline
    VDG          &            &                     &                              & \checkmark           &                    \\ \hline
    VSDG         &            &                     &                              & \checkmark (IO)      &                    \\ \hline
    \textbf{CDG} &            &                     &                              &                      & \checkmark         \\ \hline
    PDG          &            &                     &                              & \checkmark           & \checkmark         \\ \hline
    SDG          &            &                     &                              & \checkmark (inter-)  & \checkmark(inter-) \\ \hline
    SoN          &            &                     & \checkmark (Value)           & \checkmark           & \checkmark         \\ \hline
    PDW          &            &                     & \checkmark (Value+condition) & \checkmark           & \checkmark         \\ \hline
    SEG          &            &                     & \checkmark (Value)           & \checkmark (pointer) & \checkmark         \\ \hline
    \end{tabular}
    \vspace*{2mm}
    \caption{An IR syntax guide table. Each checkmark in the column indicates whether the program relationship is explicitly represented, and the text in parentheses denotes the degree of representation.}
    \label{tab:syntax}
\end{table*}

The choice of an appropriate IR syntax is the most crucial step in ensuring the subsequent analysis framework works effectively.
To make an informed decision, 
we suggest that the designer of the static analysis framework should list which program relations need to be explicitly represented and to what extent they should be represented.
Hence, we provide Tab.~\ref{tab:syntax} to assist with the decision process.
It organizes the program relations explicitly represented by each syntax discussed in this section, along with their corresponding degree of representation.

\section{IR Vocabulary}\label{sec:voc}
So far, 
we have explored the design of IR syntax,
but still need to discuss the vocabulary that fills it in to form a true IR.
Unfortunately,
past works have not directly addressed the design issue of IR vocabulary.
Therefore,
we expanded the scope of the survey to investigate various types of real-world IR designs so as to reveal the design philosophies of IR vocabulary for the readers.
In Sec.~\ref{subsec:general},
we begin with a large-scale empirical study of IR for general-purpose languages.
Next,
we perform case studies on several representative domain-specific IRs in Sec~\ref{subsec:domain}.

We break down the overall question (i.e., the design of IR vocabulary) into the following three Research Questions (RQ),
and attempt to answer them through the study.
\begin{itemize}
	\item \textbf{RQ1 (Language): } How does IR vocabulary differ among different languages?
	\item \textbf{RQ2 (Analysis): } What is special about IR vocabulary in static analysis?
	\item \textbf{RQ3 (Syntax): } Is IR vocabulary affected by the syntax?
\end{itemize}

The setting of these three questions stems from three vital considerations.
First,
one may wonder if there is a versatile vocabulary that works for all languages.
Second,
one also easily ponders that for static analysis IR, the vocabulary should be different than that in the compiler scenario.
Third,
syntax is the container for vocabulary, thus it should have some constraints on the vocabulary it carries.

\subsection{General-purpose IR}\label{subsec:general}
\textbf{Study Methodology.}
We selected the top 20 popular programming languages from TIOBE 2024 February~\cite{TIOBE}, 
an established ranking of programming languages,
as representative general-purpose IRs to examine.
The languages include \textsf{Python}, \textsf{C}, \textsf{C++}, \textsf{Java}, \textsf{C\#}, \textsf{JavaScript}, \textsf{SQL}, \textsf{Go}, \textsf{Visual Basic}, \textsf{PHP}, \textsf{Delphi}, \textsf{MATLAB}, \textsf{Assembly Language}, \textsf{Scratch}, \textsf{Swift}, \textsf{Kotlin}, \textsf{Rust}, \textsf{COBOL}, \textsf{Ruby}.
For these languages,
we consider IR from three kinds of \textbf{\textit{IR systems}},
i.e., mature compilers, static analysis research papers, and impactful industrial static analysis tools.
In particular,
an IR system may include more than one IR.
Subsequently,
we examined the \textbf{\textit{syntax}} of these IRs, 
and their basic \textbf{\textit{form of vocabulary}} under that syntax.
Additionally,
We counted whether each individual IR would highlight some typical \textbf{\textit{features}} in its vocabulary.
In other words,
the covered features can be regarded as the ``first citizen'' in the vocabulary while the uncovered ones are normally represented as library functions.
These features include control flow operations (\textsf{CF}), arithmetic computations (\textsf{Arith}), memory manipulation (\textsf{Mem}), container operations (\textsf{Array}, \textsf{Struct}, \textsf{Set}, \textsf{Map}, \textsf{Vector}), object-oriented programming (\textsf{OO}), functional programming (\textsf{FP}), module management (\textsf{Mod}), except handling (\textsf{EH}), and concurrency (\textsf{Concur}).

\begin{sidewaystable*}[]
\resizebox{\textwidth}{!}{%
\renewcommand\arraystretch{1.4}
\begin{tabular}{|l|l|l|l|l|lllllllllllll|}
\hline
\multirow{2}{*}{\textbf{IR   System Name}}                & \multirow{2}{*}{\textbf{IR Name}} & \multirow{2}{*}{\textbf{Form of Vocabulary}} & \multirow{2}{*}{\textbf{Syntax}} & \multirow{2}{*}{\textbf{Languages}}         & \multicolumn{13}{l|}{\textbf{Vocabulary   Features}}                                                                                                                                                                                                                                                                                                                                                                                        \\ \cline{6-18} 
                                                 &                          &                                     &                         &                                                                         & \multicolumn{1}{l|}{CF}         & \multicolumn{1}{l|}{Arith}      & \multicolumn{1}{l|}{Mem}        & \multicolumn{1}{l|}{Array}      & \multicolumn{1}{l|}{Struct}     & \multicolumn{1}{l|}{Map}        & \multicolumn{1}{l|}{Set}        & \multicolumn{1}{l|}{Vector}     & \multicolumn{1}{l|}{OO}         & \multicolumn{1}{l|}{FP}         & \multicolumn{1}{l|}{Mod}        & \multicolumn{1}{l|}{EH}         & Concur     \\ \hline
\multirow{2}{*}{Python   Compiler~\cite{Python}} & Python AST               & AST Node                            & AST                     & Python                                                                  & \multicolumn{1}{l|}{\checkmark} & \multicolumn{1}{l|}{\checkmark} & \multicolumn{1}{l|}{\checkmark} & \multicolumn{1}{l|}{}           & \multicolumn{1}{l|}{}           & \multicolumn{1}{l|}{\checkmark} & \multicolumn{1}{l|}{\checkmark} & \multicolumn{1}{l|}{\checkmark} & \multicolumn{1}{l|}{\checkmark} & \multicolumn{1}{l|}{\checkmark} & \multicolumn{1}{l|}{\checkmark} & \multicolumn{1}{l|}{\checkmark} & \checkmark \\ \cline{2-18} 
                                                 & Python Bytecode          & Stack                               & CFG                     & Python                                                                  & \multicolumn{1}{l|}{\checkmark} & \multicolumn{1}{l|}{\checkmark} & \multicolumn{1}{l|}{\checkmark} & \multicolumn{1}{l|}{}           & \multicolumn{1}{l|}{}           & \multicolumn{1}{l|}{\checkmark} & \multicolumn{1}{l|}{\checkmark} & \multicolumn{1}{l|}{\checkmark} & \multicolumn{1}{l|}{\checkmark} & \multicolumn{1}{l|}{\checkmark} & \multicolumn{1}{l|}{\checkmark} & \multicolumn{1}{l|}{\checkmark} & \checkmark \\ \hline
Numba~\cite{Numba}                               & Numba IR                 & Register                            & SSA                     & Python                                                                  & \multicolumn{1}{l|}{\checkmark} & \multicolumn{1}{l|}{\checkmark} & \multicolumn{1}{l|}{\checkmark} & \multicolumn{1}{l|}{}           & \multicolumn{1}{l|}{}           & \multicolumn{1}{l|}{\checkmark} & \multicolumn{1}{l|}{\checkmark} & \multicolumn{1}{l|}{\checkmark} & \multicolumn{1}{l|}{\checkmark} & \multicolumn{1}{l|}{\checkmark} & \multicolumn{1}{l|}{\checkmark} & \multicolumn{1}{l|}{\checkmark} & \checkmark \\ \hline
PyCG~\cite{PyCG}                                 & -                        & AST Node                            & AST                     & Python                                                                  & \multicolumn{1}{l|}{\checkmark} & \multicolumn{1}{l|}{\checkmark} & \multicolumn{1}{l|}{\checkmark} & \multicolumn{1}{l|}{}           & \multicolumn{1}{l|}{}           & \multicolumn{1}{l|}{}           & \multicolumn{1}{l|}{}           & \multicolumn{1}{l|}{}           & \multicolumn{1}{l|}{\checkmark} & \multicolumn{1}{l|}{\checkmark} & \multicolumn{1}{l|}{\checkmark} & \multicolumn{1}{l|}{}           &            \\ \hline
pyre-check~\cite{pyre-check}                     & -                        & AST Node                            & AST->CFG                & Python                                                                  & \multicolumn{1}{l|}{\checkmark} & \multicolumn{1}{l|}{\checkmark} & \multicolumn{1}{l|}{\checkmark} & \multicolumn{1}{l|}{}           & \multicolumn{1}{l|}{}           & \multicolumn{1}{l|}{\checkmark} & \multicolumn{1}{l|}{\checkmark} & \multicolumn{1}{l|}{\checkmark} & \multicolumn{1}{l|}{\checkmark} & \multicolumn{1}{l|}{\checkmark} & \multicolumn{1}{l|}{\checkmark} & \multicolumn{1}{l|}{\checkmark} & \checkmark \\ \hline
CIL~\cite{Necula2002CIL}                         & CIL                      & AST Node                            & AST->CFG                & C                                                                       & \multicolumn{1}{l|}{\checkmark} & \multicolumn{1}{l|}{\checkmark} & \multicolumn{1}{l|}{\checkmark} & \multicolumn{1}{l|}{\checkmark} & \multicolumn{1}{l|}{\checkmark} & \multicolumn{1}{l|}{}           & \multicolumn{1}{l|}{}           & \multicolumn{1}{l|}{}           & \multicolumn{1}{l|}{}           & \multicolumn{1}{l|}{}           & \multicolumn{1}{l|}{}           & \multicolumn{1}{l|}{}           &            \\ \hline
Clang Static Analyzer~\cite{CSA}                      & Clang AST                & AST Node                            & AST->CFG                & C/C++                                                                   & \multicolumn{1}{l|}{\checkmark} & \multicolumn{1}{l|}{\checkmark} & \multicolumn{1}{l|}{\checkmark} & \multicolumn{1}{l|}{\checkmark} & \multicolumn{1}{l|}{\checkmark} & \multicolumn{1}{l|}{}           & \multicolumn{1}{l|}{}           & \multicolumn{1}{l|}{}           & \multicolumn{1}{l|}{\checkmark} & \multicolumn{1}{l|}{}           & \multicolumn{1}{l|}{}           & \multicolumn{1}{l|}{\checkmark} &            \\ \hline
LLVM~\cite{Lattner2004LLVM}                      & LLVM IR                  & Register                            & SSA                     & C/C++~\cite{Sui2016SVF,   Schubert2019Phasar}, Go~\cite{wang2020escape} & \multicolumn{1}{l|}{\checkmark} & \multicolumn{1}{l|}{\checkmark} & \multicolumn{1}{l|}{\checkmark} & \multicolumn{1}{l|}{\checkmark} & \multicolumn{1}{l|}{\checkmark} & \multicolumn{1}{l|}{}           & \multicolumn{1}{l|}{}           & \multicolumn{1}{l|}{\checkmark} & \multicolumn{1}{l|}{}           & \multicolumn{1}{l|}{}           & \multicolumn{1}{l|}{}           & \multicolumn{1}{l|}{\checkmark} & \checkmark \\ \hline
\multirow{2}{*}{SAIL~\cite{dillig2009sail}}      & High-level IR            & AST Node                            & AST                     & \multirow{2}{*}{C}                                                      & \multicolumn{1}{l|}{\checkmark} & \multicolumn{1}{l|}{\checkmark} & \multicolumn{1}{l|}{\checkmark} & \multicolumn{1}{l|}{\checkmark} & \multicolumn{1}{l|}{\checkmark} & \multicolumn{1}{l|}{}           & \multicolumn{1}{l|}{}           & \multicolumn{1}{l|}{}           & \multicolumn{1}{l|}{}           & \multicolumn{1}{l|}{}           & \multicolumn{1}{l|}{}           & \multicolumn{1}{l|}{}           &            \\ \cline{2-4} \cline{6-18} 
                                                 & Low-level IR             & Register                            & CFG                     &                                                                         & \multicolumn{1}{l|}{\checkmark} & \multicolumn{1}{l|}{\checkmark} & \multicolumn{1}{l|}{\checkmark} & \multicolumn{1}{l|}{\checkmark} & \multicolumn{1}{l|}{\checkmark} & \multicolumn{1}{l|}{}           & \multicolumn{1}{l|}{}           & \multicolumn{1}{l|}{}           & \multicolumn{1}{l|}{}           & \multicolumn{1}{l|}{}           & \multicolumn{1}{l|}{}           & \multicolumn{1}{l|}{}           &            \\ \hline
\multirow{3}{*}{GCC~\cite{GCC}}                  & GENERIC                  & AST Node                            & AST                     & \multirow{3}{*}{C/C++}                                                  & \multicolumn{1}{l|}{\checkmark} & \multicolumn{1}{l|}{\checkmark} & \multicolumn{1}{l|}{\checkmark} & \multicolumn{1}{l|}{\checkmark} & \multicolumn{1}{l|}{\checkmark} & \multicolumn{1}{l|}{}           & \multicolumn{1}{l|}{}           & \multicolumn{1}{l|}{}           & \multicolumn{1}{l|}{}           & \multicolumn{1}{l|}{}           & \multicolumn{1}{l|}{}           & \multicolumn{1}{l|}{\checkmark} & \checkmark \\ \cline{2-4} \cline{6-18} 
                                                 & GIMPLE                   & Register                            & CFG                     &                                                                         & \multicolumn{1}{l|}{\checkmark} & \multicolumn{1}{l|}{\checkmark} & \multicolumn{1}{l|}{\checkmark} & \multicolumn{1}{l|}{\checkmark} & \multicolumn{1}{l|}{\checkmark} & \multicolumn{1}{l|}{}           & \multicolumn{1}{l|}{}           & \multicolumn{1}{l|}{}           & \multicolumn{1}{l|}{}           & \multicolumn{1}{l|}{}           & \multicolumn{1}{l|}{}           & \multicolumn{1}{l|}{\checkmark} & \checkmark \\ \cline{2-4} \cline{6-18} 
                                                 & GIMPLE SSA               & Register                            & SSA                     &                                                                         & \multicolumn{1}{l|}{\checkmark} & \multicolumn{1}{l|}{\checkmark} & \multicolumn{1}{l|}{\checkmark} & \multicolumn{1}{l|}{\checkmark} & \multicolumn{1}{l|}{\checkmark} & \multicolumn{1}{l|}{}           & \multicolumn{1}{l|}{}           & \multicolumn{1}{l|}{}           & \multicolumn{1}{l|}{}           & \multicolumn{1}{l|}{}           & \multicolumn{1}{l|}{}           & \multicolumn{1}{l|}{\checkmark} & \checkmark \\ \hline
Infer~\cite{Infer}                               & SIL                      & AST Node                            & AST->CFG                & C/C++, Java,   C\#~\cite{Infersharp}                                          & \multicolumn{1}{l|}{\checkmark} & \multicolumn{1}{l|}{\checkmark} & \multicolumn{1}{l|}{\checkmark} & \multicolumn{1}{l|}{\checkmark} & \multicolumn{1}{l|}{\checkmark} & \multicolumn{1}{l|}{}           & \multicolumn{1}{l|}{}           & \multicolumn{1}{l|}{}           & \multicolumn{1}{l|}{\checkmark} & \multicolumn{1}{l|}{}           & \multicolumn{1}{l|}{}           & \multicolumn{1}{l|}{\checkmark} &            \\ \hline
PINPOINT~\cite{Shi2018Pinpoint}                  & SEG                      & Expressions                         & PDG                     & C/C++                                                                   & \multicolumn{1}{l|}{\checkmark} & \multicolumn{1}{l|}{\checkmark} & \multicolumn{1}{l|}{\checkmark} & \multicolumn{1}{l|}{\checkmark} & \multicolumn{1}{l|}{\checkmark} & \multicolumn{1}{l|}{}           & \multicolumn{1}{l|}{}           & \multicolumn{1}{l|}{\checkmark} & \multicolumn{1}{l|}{}           & \multicolumn{1}{l|}{}           & \multicolumn{1}{l|}{}           & \multicolumn{1}{l|}{\checkmark} & \checkmark \\ \hline
Javac~\cite{Gosling1995Bytecode}                 & Java Bytecode            & Stack                               & CFG                     & Java                                                                    & \multicolumn{1}{l|}{\checkmark} & \multicolumn{1}{l|}{\checkmark} & \multicolumn{1}{l|}{\checkmark} & \multicolumn{1}{l|}{\checkmark} & \multicolumn{1}{l|}{}           & \multicolumn{1}{l|}{}           & \multicolumn{1}{l|}{}           & \multicolumn{1}{l|}{}           & \multicolumn{1}{l|}{\checkmark} & \multicolumn{1}{l|}{}           & \multicolumn{1}{l|}{}           & \multicolumn{1}{l|}{\checkmark} &            \\ \hline
LLEA~\cite{LLEA}                                 & Extended Java   Bytecode & Stack                               & CFG                     & Java, C                                                                 & \multicolumn{1}{l|}{\checkmark} & \multicolumn{1}{l|}{\checkmark} & \multicolumn{1}{l|}{\checkmark} & \multicolumn{1}{l|}{\checkmark} & \multicolumn{1}{l|}{}           & \multicolumn{1}{l|}{}           & \multicolumn{1}{l|}{}           & \multicolumn{1}{l|}{}           & \multicolumn{1}{l|}{\checkmark} & \multicolumn{1}{l|}{}           & \multicolumn{1}{l|}{}           & \multicolumn{1}{l|}{\checkmark} &            \\ \hline
\multirow{2}{*}{Soot~\cite{Vall1999Soot}}        & Jimple                   & Register                            & CFG                     & \multirow{2}{*}{Java,   Kotlin~\cite{Krishnamurthy2022}}                & \multicolumn{1}{l|}{\checkmark} & \multicolumn{1}{l|}{\checkmark} & \multicolumn{1}{l|}{\checkmark} & \multicolumn{1}{l|}{\checkmark} & \multicolumn{1}{l|}{}           & \multicolumn{1}{l|}{}           & \multicolumn{1}{l|}{}           & \multicolumn{1}{l|}{}           & \multicolumn{1}{l|}{\checkmark} & \multicolumn{1}{l|}{}           & \multicolumn{1}{l|}{}           & \multicolumn{1}{l|}{\checkmark} &            \\ \cline{2-4} \cline{6-18} 
                                                 & Shrimple                 & Register                            & SSA                     &                                                                         & \multicolumn{1}{l|}{\checkmark} & \multicolumn{1}{l|}{\checkmark} & \multicolumn{1}{l|}{\checkmark} & \multicolumn{1}{l|}{\checkmark} & \multicolumn{1}{l|}{}           & \multicolumn{1}{l|}{}           & \multicolumn{1}{l|}{}           & \multicolumn{1}{l|}{}           & \multicolumn{1}{l|}{\checkmark} & \multicolumn{1}{l|}{}           & \multicolumn{1}{l|}{}           & \multicolumn{1}{l|}{\checkmark} &            \\ \hline
Tai-e~\cite{Tan2023Tai-e}                        & -                        & Register                            & SSA                     & Java                                                                    & \multicolumn{1}{l|}{\checkmark} & \multicolumn{1}{l|}{\checkmark} & \multicolumn{1}{l|}{\checkmark} & \multicolumn{1}{l|}{\checkmark} & \multicolumn{1}{l|}{}           & \multicolumn{1}{l|}{}           & \multicolumn{1}{l|}{}           & \multicolumn{1}{l|}{}           & \multicolumn{1}{l|}{\checkmark} & \multicolumn{1}{l|}{}           & \multicolumn{1}{l|}{}           & \multicolumn{1}{l|}{\checkmark} &            \\ \hline
WALA~\cite{WALA}                                 & WALA IR                  & Register                            & SSA                     & Java,   Swift~\cite{SWAN}                                               & \multicolumn{1}{l|}{\checkmark} & \multicolumn{1}{l|}{\checkmark} & \multicolumn{1}{l|}{\checkmark} & \multicolumn{1}{l|}{\checkmark} & \multicolumn{1}{l|}{}           & \multicolumn{1}{l|}{}           & \multicolumn{1}{l|}{}           & \multicolumn{1}{l|}{}           & \multicolumn{1}{l|}{\checkmark} & \multicolumn{1}{l|}{}           & \multicolumn{1}{l|}{}           & \multicolumn{1}{l|}{\checkmark} &            \\ \hline
Graal   VM~\cite{GraalIR}                        & Graal IR                 & Expressions                         & PDG                     & Java                                                                    & \multicolumn{1}{l|}{\checkmark} & \multicolumn{1}{l|}{\checkmark} & \multicolumn{1}{l|}{\checkmark} & \multicolumn{1}{l|}{\checkmark} & \multicolumn{1}{l|}{}           & \multicolumn{1}{l|}{}           & \multicolumn{1}{l|}{}           & \multicolumn{1}{l|}{}           & \multicolumn{1}{l|}{\checkmark} & \multicolumn{1}{l|}{}           & \multicolumn{1}{l|}{}           & \multicolumn{1}{l|}{\checkmark} &            \\ \hline
OPAL~\cite{OPAL}                                 & TACAI~\cite{TACAI}       & Register                            & SSA                     & Java                                                                    & \multicolumn{1}{l|}{\checkmark} & \multicolumn{1}{l|}{\checkmark} & \multicolumn{1}{l|}{\checkmark} & \multicolumn{1}{l|}{\checkmark} & \multicolumn{1}{l|}{}           & \multicolumn{1}{l|}{}           & \multicolumn{1}{l|}{}           & \multicolumn{1}{l|}{}           & \multicolumn{1}{l|}{\checkmark} & \multicolumn{1}{l|}{}           & \multicolumn{1}{l|}{}           & \multicolumn{1}{l|}{\checkmark} &            \\ \hline
MSIL~\cite{ECMA-335}                             & MSIL                     & Register                            & CFG                     & C\#, Visual Basic                                                                     & \multicolumn{1}{l|}{\checkmark} & \multicolumn{1}{l|}{\checkmark} & \multicolumn{1}{l|}{\checkmark} & \multicolumn{1}{l|}{\checkmark} & \multicolumn{1}{l|}{\checkmark} & \multicolumn{1}{l|}{}           & \multicolumn{1}{l|}{}           & \multicolumn{1}{l|}{}           & \multicolumn{1}{l|}{\checkmark} & \multicolumn{1}{l|}{}           & \multicolumn{1}{l|}{}           & \multicolumn{1}{l|}{\checkmark} &            \\ \hline
JSAI~\cite{JSAI}                                 & notJS                    & AST Node                            & AST->CFG                & JavaScript                                                              & \multicolumn{1}{l|}{\checkmark} & \multicolumn{1}{l|}{\checkmark} & \multicolumn{1}{l|}{\checkmark} & \multicolumn{1}{l|}{}           & \multicolumn{1}{l|}{}           & \multicolumn{1}{l|}{}           & \multicolumn{1}{l|}{}           & \multicolumn{1}{l|}{}           & \multicolumn{1}{l|}{\checkmark} & \multicolumn{1}{l|}{}           & \multicolumn{1}{l|}{}           & \multicolumn{1}{l|}{\checkmark} &            \\ \hline
Hauzer et   al.~\cite{hauzar2015framework}       & -                        & AST Node                            & AST->CFG                & PHP                                                                     & \multicolumn{1}{l|}{\checkmark} & \multicolumn{1}{l|}{\checkmark} & \multicolumn{1}{l|}{\checkmark} & \multicolumn{1}{l|}{}           & \multicolumn{1}{l|}{}           & \multicolumn{1}{l|}{}           & \multicolumn{1}{l|}{}           & \multicolumn{1}{l|}{}           & \multicolumn{1}{l|}{\checkmark} & \multicolumn{1}{l|}{}           & \multicolumn{1}{l|}{}           & \multicolumn{1}{l|}{\checkmark} &            \\ \hline
HHVM~\cite{HHVM}                                 & HHIR~\cite{HHIR}         & Register                            & SSA                     & PHP                                                                     & \multicolumn{1}{l|}{\checkmark} & \multicolumn{1}{l|}{\checkmark} & \multicolumn{1}{l|}{\checkmark} & \multicolumn{1}{l|}{\checkmark} & \multicolumn{1}{l|}{}           & \multicolumn{1}{l|}{\checkmark} & \multicolumn{1}{l|}{\checkmark} & \multicolumn{1}{l|}{\checkmark} & \multicolumn{1}{l|}{\checkmark} & \multicolumn{1}{l|}{\checkmark} & \multicolumn{1}{l|}{\checkmark} & \multicolumn{1}{l|}{\checkmark} & \checkmark \\ \hline
Bastet~\cite{Bastet}                             & LelLa                    & AST Node                            & AST->CFG                & Scratch                                                                 & \multicolumn{1}{l|}{\checkmark} & \multicolumn{1}{l|}{\checkmark} & \multicolumn{1}{l|}{\checkmark} & \multicolumn{1}{l|}{}           & \multicolumn{1}{l|}{}           & \multicolumn{1}{l|}{}           & \multicolumn{1}{l|}{}           & \multicolumn{1}{l|}{\checkmark} & \multicolumn{1}{l|}{}           & \multicolumn{1}{l|}{}           & \multicolumn{1}{l|}{}           & \multicolumn{1}{l|}{}           & \checkmark \\ \hline
GoSSA~\cite{GoSSA}                               & GoSSA                    & Register                            & SSA                     & Go~\cite{Lange2018Go,   GCatch}                                     & \multicolumn{1}{l|}{\checkmark} & \multicolumn{1}{l|}{\checkmark} & \multicolumn{1}{l|}{\checkmark} & \multicolumn{1}{l|}{\checkmark} & \multicolumn{1}{l|}{\checkmark} & \multicolumn{1}{l|}{\checkmark} & \multicolumn{1}{l|}{}           & \multicolumn{1}{l|}{}           & \multicolumn{1}{l|}{}           & \multicolumn{1}{l|}{\checkmark} & \multicolumn{1}{l|}{}           & \multicolumn{1}{l|}{\checkmark} & \checkmark \\ \hline
Flang   Compiler~\cite{FortranIR}                & FortranIR                & Register                            & SSA                     & Fortran                                                                 & \multicolumn{1}{l|}{\checkmark} & \multicolumn{1}{l|}{\checkmark} & \multicolumn{1}{l|}{\checkmark} & \multicolumn{1}{l|}{\checkmark} & \multicolumn{1}{l|}{}           & \multicolumn{1}{l|}{}           & \multicolumn{1}{l|}{}           & \multicolumn{1}{l|}{}           & \multicolumn{1}{l|}{}           & \multicolumn{1}{l|}{}           & \multicolumn{1}{l|}{}           & \multicolumn{1}{l|}{}           &            \\ \hline
McSAF~\cite{McSAF}                               & MCLAST                   & AST Node                            & AST->CFG                & MATLAB                                                                  & \multicolumn{1}{l|}{\checkmark} & \multicolumn{1}{l|}{\checkmark} & \multicolumn{1}{l|}{\checkmark} & \multicolumn{1}{l|}{\checkmark} & \multicolumn{1}{l|}{}           & \multicolumn{1}{l|}{}           & \multicolumn{1}{l|}{}           & \multicolumn{1}{l|}{}           & \multicolumn{1}{l|}{}           & \multicolumn{1}{l|}{\checkmark} & \multicolumn{1}{l|}{}           & \multicolumn{1}{l|}{}           &            \\ \hline
Swift   Compiler~\cite{SIL}                      & SIL                      & Register                            & SSA                     & Swift                                                                   & \multicolumn{1}{l|}{\checkmark} & \multicolumn{1}{l|}{\checkmark} & \multicolumn{1}{l|}{\checkmark} & \multicolumn{1}{l|}{\checkmark} & \multicolumn{1}{l|}{\checkmark} & \multicolumn{1}{l|}{}           & \multicolumn{1}{l|}{}           & \multicolumn{1}{l|}{}           & \multicolumn{1}{l|}{\checkmark} & \multicolumn{1}{l|}{}           & \multicolumn{1}{l|}{}           & \multicolumn{1}{l|}{}           &            \\ \hline
RIL~\cite{RIL}                                   & RIL                      & AST Node                            & AST                     & Ruby~\cite{DRuby}                                                       & \multicolumn{1}{l|}{\checkmark} & \multicolumn{1}{l|}{\checkmark} & \multicolumn{1}{l|}{\checkmark} & \multicolumn{1}{l|}{\checkmark} & \multicolumn{1}{l|}{}           & \multicolumn{1}{l|}{}           & \multicolumn{1}{l|}{}           & \multicolumn{1}{l|}{}           & \multicolumn{1}{l|}{\checkmark} & \multicolumn{1}{l|}{}           & \multicolumn{1}{l|}{}           & \multicolumn{1}{l|}{}           &            \\ \hline
PowerStation~\cite{PowerStation}                 & -                        & Expressions                         & PDG                     & Ruby                                                                    & \multicolumn{1}{l|}{\checkmark} & \multicolumn{1}{l|}{\checkmark} & \multicolumn{1}{l|}{\checkmark} & \multicolumn{1}{l|}{\checkmark} & \multicolumn{1}{l|}{}           & \multicolumn{1}{l|}{}           & \multicolumn{1}{l|}{}           & \multicolumn{1}{l|}{}           & \multicolumn{1}{l|}{\checkmark} & \multicolumn{1}{l|}{}           & \multicolumn{1}{l|}{}           & \multicolumn{1}{l|}{\checkmark} & \checkmark \\ \hline
\multirow{2}{*}{RustC   Compiler~\cite{Rust}}    & HIR                      & AST Node                            & AST                     & Rust                                                                    & \multicolumn{1}{l|}{\checkmark} & \multicolumn{1}{l|}{\checkmark} & \multicolumn{1}{l|}{\checkmark} & \multicolumn{1}{l|}{\checkmark} & \multicolumn{1}{l|}{\checkmark} & \multicolumn{1}{l|}{}           & \multicolumn{1}{l|}{}           & \multicolumn{1}{l|}{}           & \multicolumn{1}{l|}{}           & \multicolumn{1}{l|}{}           & \multicolumn{1}{l|}{}           & \multicolumn{1}{l|}{}           &            \\ \cline{2-18} 
                                                 & MIR                      & Register                            & CFG                     & Rust~\cite{MirChecker,   SafeDrop}                                      & \multicolumn{1}{l|}{\checkmark} & \multicolumn{1}{l|}{\checkmark} & \multicolumn{1}{l|}{\checkmark} & \multicolumn{1}{l|}{\checkmark} & \multicolumn{1}{l|}{\checkmark} & \multicolumn{1}{l|}{}           & \multicolumn{1}{l|}{}           & \multicolumn{1}{l|}{}           & \multicolumn{1}{l|}{}           & \multicolumn{1}{l|}{}           & \multicolumn{1}{l|}{}           & \multicolumn{1}{l|}{}           &            \\ \hline
\end{tabular}%
}
\vspace*{1mm}
\caption{A study of IR vocabulary in real-world IRs}
\label{tab:voc}
\end{sidewaystable*}
\noindent\textbf{Results.}
Tab.~\ref{tab:voc} summarizes the main study results.
Overall,
we have identified 29 IR systems that encompass 16 out of the 20 languages.
SQL and Assembly language are not applicable to the concept of IR. No relevant information was found for COBOL and Delphi.
Additionally,
we have found that some impactful static analysis studies rely on the IR of an existing IR system.
For such cases,
we would cite the study in the corresponding languages in the \textit{Languages} column.
Next,
we delve into these data to examine each of the research questions.

\subsubsection{RQ1 (Language)}
\begin{finding}\label{find:lang1}
\itshape
The vocabulary of IR basically represents a low-level mapping of the fundamental features of the language.
\end{finding}
By comparing the coverage of different IR systems of the same language in column \textit{Vocabulary Features},
it could be revealed that the supported features largely align with the language's inherent capabilities,
therefore are very comparable.
For instance,
in the case of C/C++'s IRs, 
there is a strong emphasis on structural elements and array operations.
Similarly, 
Java's IRs prominently support object-oriented programming, array operations, and error handling.
Another example is Python, 
which offers a wide range of features at the language level, 
including various containers such as dictionaries, sets, and lists, as well as functional programming and module management. 
This necessitates that its IR also supports these features.

Furthermore,
some IR systems also accommodate specific non-built-in yet widely utilized features.
LLVM IR~\cite{Lattner2004LLVM} includes atomic operations to facilitate the representation of different concurrency models.
Similarly, 
GIMPLE IR~\cite{GCC} contains several concurrency primitives that are tailored to the OpenMP library.
Hence,
the design principles of language features are presumably influenced by economic considerations.
Specifically,
commonly utilized features include a higher frequency of real-world programs.
Considering them as first-class citizens in the IR and prioritizing their handling in static analyzers is a more economical choice.

\begin{finding}\label{find:lang2}
\itshape
The substantial variation in features across different languages results in significant differences in the design of IR vocabulary.
\end{finding}
The disparities in IR vocabulary among different languages are very evident when compared vertically.
Only basic operations like \textsf{CF}, \textsf{Arith}, and \textsf{Mem} are common among various IRs.
Notably,
instances of the same features indicated in Tab.~\ref{tab:voc} can vary between various IRs.
For example, the IRs of Python and Java conform to their own object-oriented programming standards.
Go's IR,
GoSSA~\cite{GoSSA} uses message passing for concurrency operations,
whereas IRs for C/C++ rely on shared memory for concurrency operations.

There are some IRs that can represent multiple languages.
Initially IRs like LLVM IR, GIMPLE, MSIL~\cite{ECMA-335}, and Jimple~\cite{Vall1999Soot} have a relatively low level of abstraction with respect to the languages they represent, 
allowing them to represent multiple languages using a limited set of vocabulary.
However, this method is only suitable for languages that share many similarities and loses many advanced language features.
Additionally, the LLEA framework~\cite{LLEA} utilizes Java Bytecode to depict C programs for analyzing Java Native Interfaces (JNI).
This approach is advantageous since Java IRs are more expressive compared to IRs for C.
Similarly,
C++ IRs can effectively represent C programs.
Thirdly, IRs such as HHIR~\cite{HHIR} and SIL~\cite{Infer} expand their vocabulary to incorporate more language features.
For instance,
the Smallfoot Intermediate Language (SIL) of Infer was first tailored for C/C++, 
but later enhanced to incorporate features of Java and C\# through the integration of language-specific vocabularies.
This approach's downside is the inflation of IR vocabulary, rendering IR itself incomprehensible.

\subsubsection{RQ2 (Analysis)}
\begin{finding}\label{find:analysis}
\itshape
IR for static analysis can be designed in a more simplified and less ambiguous manner.
\end{finding}

Compared with IR utilized in compilers,
we identify that IR for static analysis possesses two characteristics.
First,
static analysis IR uses simpler vocabulary.
Specifically,
CIL~\cite{Necula2002CIL} removes syntactic sugars in C like \textsf{->} and lifts all the type declarations to the global scope.
Additionally,
Jimple merges different types of \textsf{invoke} instructions in Java Bytecode,
such as \textsf{InvokeVirtual}, \textsf{InvokeSpecial}, and \textsf{InvokeStatic},
to provide a general interface to retrieve the callee function.
Second,
static analysis IR carries less ambiguity in the vocabulary.
Both CIL and SAIL~\cite{dillig2009sail} aim to clarify the \textit{l-value} in C.
For example, a statement like \textsf{A[10]} is represented differently depending on whether \textsf{A} is a pointer array or not.
Furthermore, the IR notJS~\cite{JSAI} ensures that its AST expressions consist of pure computation that always terminates without exception.
Moreover,
Tai-e~\cite{Tan2023Tai-e} explicitly distinguishes different assignment statements, 
allowing developers to know the type of expression on the right-hand side explicitly.
These skills make the designed IR vocabulary more friendly to static analysis.

\subsubsection{RQ3 (Syntax)}
\begin{finding}\label{find:syntax}
\itshape
The four syntaxes, AST, CFG, SSA, and PDG are most popular in real-world IRs.
Specifically,
AST and CFG serve as a boundary that affects the abstraction level of the IR vocabulary.
\end{finding}

All four core syntaxes (AST, CFG, SSA, PDG) introduced in Sec.~\ref{sec:syntax} are used in real-world IRs.
The difficulty of constructing these four syntaxes gradually increases. 
Moreover, there are progressive relationships among their constructions.
Thus, the universality diminishes progressively from AST to PDG based on the number of syntaxes.
Moreover,
there are four main forms of vocabulary.
The AST node is applied under AST.
CFG mainly employs register-based instructions and stack-based instructions.
However,
it is again confirmed that the latter is not appropriate for static analysis.
For instance,
static analysis of Java usually starts by first converting Java Bytecode~\cite{Gosling1995Bytecode} to register-based IR, 
such as Jimple.
In addition,
although the IR of LLEA is stack-based, 
the subsequent utilization of Jlint in LLEA still recovers mutual data dependencies between instructions,
which is equivalent to transforming them into register instructions.
SSA, which further emphasizes data dependence,
is often based on register instructions.
Finally,
the basic vocabulary form of PDG is expressions.

Additionally, 
it can be observed that there is an abstract level difference between AST and CFG.
AST allows for scoping, 
which enables the representation of certain features in a context-specific manner, 
such as the \textsf{try-catch} block in Java or the \textsf{with} statement in Python.
On the CFG, 
such features need to be implemented collectively through several instructions.
To analyze them, 
it is necessary to establish a state machine based on control flow and data flow between instructions.
Observing such differences in abstraction levels,
the design of SAIL~\cite{dillig2009sail} includes both high-level and low-level IR, 
establishing a one-to-many mapping relationship.
Thanks to this design, 
SAIL utilizes two abstraction levels to describe static analysis tasks, 
such as \textit{``Can the lack of a break statement in some branch of a switch construct lead to an error?''}

\remark{
\textbf{Lessons learned in general-purpose IR.}
Based on Finding~\ref{find:lang1} and \ref{find:lang2}, 
we believe there is no one-size-fits-all vocabulary for various languages.
Basically, the design of vocabulary should capture the core features of the target language (language family).
According to Finding~\ref{find:analysis},
static analysis imposes specific requirements on the vocabulary as well,
leading to some vocabulary being simplified,
and some being elaborated upon.
Finding~\ref{find:syntax} highlights the impact of IR syntax on the abstraction level of vocabulary. 
Therefore, when designing an IR, 
we need to consider which abstraction level is most suitable for analyzing the target language features and choose the syntax accordingly.
}

\subsection{Domain-specific IR}\label{subsec:domain}
Over the past few years,
we have witnessed the emergence of many domain-specific computations,
which have posed new requirements for the design of IR.
To further answer these research questions,
we conduct case studies on IR designs in three representative domains (\S~\ref{subsubsec:Parallel}, \ref{subsubsec:HD}, \ref{subsubsec:ML}) and introduce an influential framework MLIR for creating new domain-specific IRs (\S~\ref{subsubsec:MLIR}).

\subsubsection{IR for Parallel Computing}\label{subsubsec:Parallel}
Parallel computing decomposes a computational task into subtasks and distributes them to different processing units such as processors, nodes, or accelerators.
The IR of parallel programs needs to explicitly represent the synchronization between computation units in different parallelism mechanisms~\cite{Susungi2021IRParallel}.

To achieve this,
the first lines of works~\cite{Schardl2019Tapir, Zhao2011PIR, Khaldi2013SPIREA} focus on expanding parallel primitives in the operation (instruction) set of IR.
To support fork-join parallelism, where a parent routine spawns several subroutines in parallel,
Tapir~\cite{Schardl2019Tapir} extends LLVM instruction set with three instructions, \textsf{detach}, \textsf{reattach}, and \textsf{sync}.
In order to better support task parallelism,
PIR~\cite{Zhao2011PIR} transforms the high-level constructs in the Habanero-Java library into mid-level IR. 
This is achieved through a combination of basic instructions and three parallel instructions, namely \textsf{async}, \textsf{finish}, and \textsf{isolated}.
 To express the message-passing model of the distributed memory system,
 SPIRE~\cite{Khaldi2013SPIREA} extends PIPS~\cite{Irigoin1991PIPS} and LLVM using four instructions, \textsf{send}, \textsf{recv}, and \textsf{signal}, \textsf{wait}.
 
 Moreover, 
 in order to distinguish between parallel and serial control flow, another group of works~\cite{Lee1999CCFG, Sarkar1993PPG, Srinivasan1992EFG} extends the definition of nodes and edges in the CFG to incorporate parallelism.
Firstly, to represent the creation of a new thread,
a block (node) in CFG is no longer limited to a single instance, but can be spawned.
For example, Parallel Program Graph (PPG~\cite{Sarkar1993PPG}) defines a \textsf{MGOTO} block, 
which ends with a thread creation instruction. 
When the control flow leaves this block, 
each of its successor blocks will be spawned,
corresponding to the newly created threads that will execute in parallel.
Meanwhile,
Extended Flow Graph (EFG~\cite{Srinivasan1992EFG}) uses region to group related parallel blocks together, 
making it a subgraph in the sequential CFG,
which narrows down the scope of analysis.
Secondly,
to express the synchronization between threads,
the definition of CFG edges has been expanded from purely control flow transitions to support synchronization.
In the design of Concurrent Control Flow Graph~\cite{Lee1999CCFG}(CCFG~\cite{Lee1999CCFG}),
two types of edge are incorporated: \textsf{conflict} edges and \textsf{synchronization} edges.
In particular,
a block that ends with an instruction that releases a resource can point to another block starting with an instruction that acquires the same resource through a \textsf{conflict} edge.
Likewise, a block that ends with an instruction that signals another thread can point to another block starting with an instruction that waits for the signal through a \textsf{synchronization} edge.
These works reflect a broader vocabulary design,
encompassing more than just operations.

\subsubsection{IR for Hardware Design} \label{subsubsec:HD}
Hardware Description Languages (HDLs) are utilized to specify the high-level behavior and structure of digital circuits.
Designing IR for HDLs presents significant challenges in terms of abstracting the elements of digital circuits and their associated behaviors.

A new IR LLHD~\cite{Schuiki2020LLHD} is introduced to address these challenges.
Specifically,
it supports the memory model, operations, and modular abstraction for hardware.
Firstly,
in the memory model,
it involves three core hardware concepts,
\textsf{signal}, \textsf{register}, and \textsf{memory},
associated with timing control, instruction execution, and data storage.
Secondly,
it extends the common IR instruction set with timing control instructions,
such as the \textsf{wait} instruction, 
which halts program execution until a signal change is received.
Third,
three abstractions, \textsf{functions}, \textsf{processes}, and \textsf{entities},
are provided to support hardware modularity.
\textsf{Functions} correspond to combinational logic circuits, 
capturing pure function mappings from input values to output values. 
\textsf{Processes} correspond to sequential logic circuits, 
which utilize intermediate states to respond to the order and timing of inputs with outputs.
\textsf{Entities} correspond to sub-modules in digital circuits, 
constructing larger hierarchical structures by instantiating other functions, processes, or entities.
These can be seen as the extension of function abstraction in general IRs.

\subsubsection{IR for Machine Learning}\label{subsubsec:ML}
Machine learning (ML) languages~\cite{PyTorch, Tensorflow} are designed to facilitate the development, training, and deployment of various ML models in distributed, parallel, and heterogeneous computing environments.
The design focus of IR is to better represent the ML models,
thereby understanding their inner operations.

Glow~\cite{Rotem2018GlowGL} proposes using the computational graph as IR,
enabling the representation of different neural networks (NN).
Specifically,
a computational graph consists of interconnected neurons (nodes) and numerous parameters encoded on the edges.
Based on this abstraction,
Glow is capable of performing constant propagation and quantization.
The former eliminates redundant computations by propagating constant values such as pre-trained weights within the graph,
while the latter converts heavy-weighted floating-point operations to light-weighted low-bit integer operations by analyzing the value range of neuron outputs.
Therefore, fundamentally, the computational graph can be seen as the data dependence graph on the specialized IR values, i.e. neurons.
Moreover,
Relay~\cite{Roesch2018Relay} addresses the prevalent differentiate computation in DL.
It extends the computational graph with a \textsf{Grad} operator,
presenting the calculation of the gradient of a higher-order function.
This operator is useful for computing parameter gradients with respect to the output during model training.

%

\subsubsection{Multi Level IR Design}\label{subsubsec:MLIR}
As demonstrated in the previous cases,
designing a new IR has become a common problem in many domains.
Therefore,
the Multi-level Intermediate Representation (MLIR~\cite{Lattner2021MLIR}) is introduced as a framework to facilitate this process.
To represent a new computation,
MLIR allows users to create domain-specific IR by defining a new \textsf{dialect},
which contains new data types, operations, and attributes, as well as infrastructures such as analyses, optimizations, and transformations,
as a typical IR.
Another new concept introduced in MLIR is \textsf{region}, 
which organizes sequences of operations and supports the nesting of other regions. 
By combining operations and regions in the IR, 
the higher-level computation models and control structures can be represented,
such as for loops or a layer in a computational graph.

MLIR introduces a technique named partial lowering.
Specifically,
different levels of IR operations can coexist,
and high-level operations can gradually be transformed into lower-level operations.
In this process,
as different high-level operations may be transformed into several common low-level operations,
the modules corresponding to low-level operations can be reused in different domain-specific IRs,
greatly increasing maintainability.
To enhance the efficiency of IR design,
MLIR further adopts the use of domain-specific languages.
Through IR declaration tools like TableGen and IRDL~\cite{Fehr2022IRDL},
the tedious implementation details can be automated.

MLIR has already been instrumental in designing domain-specific IRs for a variety of areas, such as hardware design~\cite{majumder2021hir}, functional programming~\cite{CC2021FP}, and quantum computing~\cite{nguyen2021quantum}.

\subsubsection{Answer to Research Questions}\label{subsubsec:answer-domain}
After studying the design of the domain-specific IRs,
we proceed to address the research questions to provide additional insights.

\begin{finding}
\textit{
In domain-specific IR,
the level of abstraction in the IR vocabulary is higher.
They are used to represent diverse computational concepts in specific domains.
}
\end{finding}

Regarding RQ1, findings~\ref{find:lang1} and \ref{find:lang2} still hold.
However,
when representing the computational nature in various domains,
there is a substantial increase in the abstraction level of IR vocabulary.
Notably,
the increment in vocabulary in domain-specific IRs compared to general-purpose IRs can touch various elements within IR,
such as new IR instructions/operations~\cite{Schardl2019Tapir}, new types of control flow edges~\cite{Lee1999CCFG}, new modular abstractions~\cite{Schuiki2020LLHD}, and new IR values~\cite{Rotem2018GlowGL}.
This highlights the high degree of freedom in designing the IR vocabulary,
serving as a supporting point for the design of operations and regions in MLIR. 
Consequently,
the overlap of IR vocabulary between domain-specific IRs will be significantly reduced compared to general-purpose IRs.
However,
it is worth noting that during the IR lowering process, 
the vocabulary overlap gradually increases again,
providing opportunities for modular reuse in low-level IR.

With regard to RQ2, the community lacks research demonstrating static analysis in the context of domain-specific IR.
Therefore,
we discuss this topic as future works in \S~\ref{sec:discussion}.

\begin{finding}
\textit{
In different domains,
the essence of IR syntax remains fundamentally unchanged,
but it gives birth to numerous instances specific to those domains.
}
\end{finding}

For RQ3,
although it may seem that these domain-specific IRs have introduced new graph representations,
they can be regarded as higher-level instances of classical IR syntaxes, 
i.e. program relationships.
For example,
PPG~\cite{Sarkar1993PPG}, EFG~\cite{Srinivasan1992EFG}, CCFG~\cite{Lee1999CCFG} are all extensions of control flow relationships in parallel scenarios,
and computational graphs~\cite{Roesch2018Relay,Rotem2018GlowGL} represent the data dependency relationships between neurons.
Moreover,
it can be observed that IR syntax and abstraction levels are not inherently tied.
For example, the programming model in machine learning determines that a dependence graph can serve as the syntax of the high-level IR. 
However, when considering low-level implementation details, 
we need to rely on IR syntax like AST or CFG to establish the low-level IR. 
Therefore, the intuition that AST to CFG to PDG progressively lowers in a general-purpose IR does not apply here.

\remark{
\textbf{Lessons learned in domain-specific IR.}
In order to better represent the computations in a specific domain, 
we can flexibly expand different kinds of IR elements and consider the unique manifestations of program relationships involved in these computations,
in order to reconcile the syntax and vocabulary of high-level IR.
}

\section{IR Query}\label{sec:query}
\begin{table*}[t]
\centering
\renewcommand\arraystretch{1.4}
\begin{tabular}{|l|l|l|}
\hline
\textbf{No.} & \textbf{Analysis Task}                                                                                                                                      & \textbf{Summary}       \\ \hline
\textbf{I}   & \begin{tabular}[c]{@{}l@{}} On AST, find all call sites where the callee's name starts with keyword \\ ``mem'' and the arguments are all constant\end{tabular}   & Basic Pattern Matching \\ \hline
\textbf{II}  & \begin{tabular}[c]{@{}l@{}} On CFG, check whether each Java \textsf{FileInputStream} object is both read \\ at least once and properly closed in the end.\end{tabular} & Typestate property              \\ \hline
\textbf{III} & \begin{tabular}[c]{@{}l@{}} On ICFG, check whether function \textsf{foo()} is reachable from main function\end{tabular}                                            & Call Stack Pattern \\ \hline
\textbf{IV}  & Implement live variable analysis, reaching definition analysis, \dots                                                                                         & Various Dataflow Analyses      \\ \hline
\textbf{V}   & Detect SQL injection, Cross-Site Scripting, \dots                                                                                                            & Various Taint Analyses         \\ \hline
\end{tabular}
\caption{Example of different analysis tasks}
\label{tab:analysis-tasks}
\end{table*}

Through the design of IR syntax and vocabulary,
the program information has been sufficiently provided.
Nevertheless,
it is far from sufficient only to use the basic functionalities of the IR to accomplish a wide range of analyses.
Consider implementing the five analysis tasks listed in Tab.~\ref{tab:analysis-tasks},
where an adequate IR has already been provided.
Evidently, 
it remains non-trivial to acquire the ``desired solution'' from the ``search space'',
i.e. obtaining an effective and efficient implementation using the IR's bare APIs.
Therefore,
to fully unleash the power of the designed IR,
we should encapsulate the fundamental program information provided by the IR into a query library.
This topic encompasses a wide range of studies, 
and previous researchers mostly focus on designing the query system from two perspectives.
Firstly, they further enhance the advantages of the IR
in order to accommodate higher-level program relationships (\S~\ref{subsec:IR-query}).
This benefits tasks such as I-III in Tab.~\ref{tab:analysis-tasks}.
Secondly,
their attention lies on serving specific analysis purposes (\S~\ref{subsec:Analysis-query}),
facilitating tasks such as IV and V in Tab.~\ref{tab:analysis-tasks}.

\subsection{IR-centered Query}\label{subsec:IR-query}

\subsubsection{Vocabulary}
Consider task I in Tab.~\ref{tab:analysis-tasks}, which only concerns a single node in AST.
However,
a developer need to precisely specify its kind information and attribute information to filter it out from all the nodes with various kinds and attributes.
Some studies aim to simplify such queries.

To support querying kind information,
JQuery~\cite{JQuery} uses bool predicates such as $\textsf{package}(e)$, $\textsf{type}(e)$, and $\textsf{class}(e)$ to check an IR element $e$ is of certain kind.
Similarly,
ASTLOG~\cite{Crew1997ASTLOG} utilizes enumeration values,
referred to as \textit{opcode},
to record different kind of IR elements.
Then,
the analysis logic related to various elements can be dispatched based on this opcode.
For the property related queries,
regular expressions are intensively used to describe the string patterns such as the names of types, functions, and so on~\cite{Crew1997ASTLOG, Zhang2003Aspects}.
Additionally Prism~\cite{Zhang2003Aspects} inspired by AspectJ~\cite{AspectJ} supports the use of wildcard $*$ to describe the concept of arbitrariness when matching the type signatures.

\subsubsection{Syntax}
Compared to IR vocabulary,
IR syntax has received more attention from researchers,
due to the fact that for different program relationships,
there should be different mechanisms to organize the query.
Notably, the current focus revolves around the three core syntaxes presented in \S~\ref{sec:syntax}, i.e. AST, CFG, and PDG.
There is no explicit effort to improve SSA's query capabilities, 
given SSA is a variant of CFG itself.
In addition, the def-use chain introduced by SSA is efficiently convenient.

\textit{\textbf{AST.}} Analyses on AST can be considered as the task of finding the desired subtrees,
which can then be achieved through tree traversals.
The optimization system OPTRAN~\cite{lipps1989optran} design a two phase AST analysis,
which involves first traversing each AST node to calculate and propagate some invariants and then matching the subtrees in a pre-order by utilizing these invariants.
The Miniphase compiler~\cite{PetrashkoLO17} designs each of its passes as a visitor of tree nodes so that separate passes can be fused together and performed within a single post-order traversal of AST.
Moreover,
Tree traversal synthesizers like Hecate~\cite{Chen2022tree} have designed their query languages based on attribute grammar~\cite{attribute-grammar}.
Specifically,
a tree node is defined as a class with multiple children nodes and attributes that can be primitive or recursive.
Subsequently,
a tree traversal matches the types of tree nodes and either accesses their corresponding attributes or recursively traverses their child nodes for specific tree nodes.
However, 
attribute grammar cannot express AST traversals such as constant propagation,
thus some works on tree traversal fusion, 
such as TreeFuser~\cite{Sakka2017recursive} and Grafter~\cite{Sakka2019pldi} improve the traversal language, such as allowing the children node to have any type.

\textit{\textbf{CFG.}}
Given the nature of basic blocks,
an intuitive query requirement on CFG is a continuous sequence of instructions,
which often expresses a piece of localized computation.
Specifically,
a typical peephole optimizer~\cite{Peephole, Davidson1980Peephole} would match instruction sequences using a sliding window.
However,
such a scheme is incapable of addressing queries about the type-state properties within the execution paths, 
i.e. task II in Tab.~\ref{tab:analysis-tasks} or use-after-free bug.
PRIME, an API-usage checking tool, achieves this goal by designing queries as a sequence of operations on multiple objects, with unknown operations used as wildcards. 
To achieve this objective, 
PRIME~\cite{PRIME}, 
a tool designed to check API usage, constructs queries as a sequence of operations performed on multiple Java objects.
This query also permits the use of "unknown" operations as wildcards.
Subsequently, the query is transformed into a finite state machine to match API usage patterns in CFG. 
Many studies on program transformations~\cite{Lacey2001Rewrite, Kalvala2009Temporal} expand such queries by employing temporal logic side conditions to articulate the matching condition.
Another study, EXPLORER~\cite{Feng2015EXPLORER}, aims to better express inter-procedural control flow features using ICFG.
The query is formulated using regular expressions with arrows to identify a potential call stack pattern.
For example, 
task III in Tab.~\ref{tab:analysis-tasks} can be achieved via a query \textsf{main $\to$ .* $\to$ foo} denotes the transitive reachability from the \textsf{main} function to the \textsf{foo} function.
EXPLORER have proven to be useful in different scenarios, such as detecting design patterns,  performance issues, and cross-package invocations.

\textit{\textbf{PDG.}}
A query language DQL~\cite{Wang2010MatchSDG} is proposed to unleash the power of SDG.
DQL enables users to describe a program point by specifying several constraints on it related to data dependence, control dependence, and call relationships.
PIDGIN~\cite{Johnson2015SecPDG},
an information security analyzer, 
leverages the strong connection between PDG and program paths. 
It has introduced a query language with a high level of expressiveness for constructing paths.
More precisely, PIDGIN receives a query in the form of a path description obtained by either forward or backward slicing of a PDG node, 
or by considering two source-sink nodes that may be connected.
In addition, it allows for path queries by temporarily excluding specific nodes or dependency edges on the PDG, 
which is highly useful for conducting sanitizer-related information flow queries.

\textit{\textbf{Multiple IR.}}
Realizing that specific analysis may depend on the cooperation of multiple IR syntaxes,
Yamaguchi et al.~\cite{Yamaguchi2014CPG} introduced Code Property Graph (CPG) which combines AST, CFG, and PDG.
It is designed as a multi-graph,
denoted as $(V = V_{A}, E = E_{A} \cup E_{C} \cup E_{DD} \cup E_{CD})$.
Notably,
all the nodes originate from AST nodes $V_{A}$.
To achieve a unified representation,
the program relation edges are computed solely from AST nodes,
but trade-offs are made to reduce the explicitness of program relations.
Specifically,
The control flow edges $E_{C}$ and control dependence edges $E_{CD}$ are established on the AST nodes of the statement and predicate,
instead of basic blocks.
The data dependence edges $E_{DD}$ are established on the variable level.
This compromise, 
however, 
provides flexibility to describe various security vulnerabilities using multi-graph traversals.
\begin{itemize}
\item 
\textit{Syntax-only vulnerabilities} are identified by matching on AST nodes $V_A$,
specifically by checking if the names and types meet certain criteria during the traversal of $V_A$ with no particular order.
It is worth noting that a single vulnerability can include multiple matches.
It handles vulnerabilities such as insecure arguments. 
\item
\textit{Control-flow vulnerabilities} involve matching all program paths from some start nodes to some end nodes. 
In each path, a traversal is conducted through the control flow edges, 
examining multiple nodes along the way to infer the safety of that particular path.
Vulnerabilities such as resource leaks and use-after-free can be handled.
\item
\textit{Taint-style vulnerabilities} are detected by initially locating source-sink node pairs through traversals using data dependence edges $E_{DD}$.
Subsequently,
the paths between the identified source-sink pairs are scrutinized to determine the presence of sanitization checks.
\end{itemize}

\subsection{Analysis-centered Query}\label{subsec:Analysis-query}
Some research contends that the query should be designated to facilitate the implementation of targeted static analysis. 
Consequently, the analysis writers could focus on the fundamental concepts associated with the analysis.
Nevertheless, it is the query designer's responsibility to build the relationship between these concepts and the underlying IR.

\textit{\textbf{Dataflow Analysis.}}
Back to task IV in Tab.~\ref{tab:analysis-tasks}.
A typical dataflow analysis iterates to compute flow values based on different flow functions.
Sharlit~\cite{Tjiang1992Sharlit},
a compilation framework,
provides a flexible approach to implement dataflow analysis.
To establish dataflow analysis in Sharlit,
one first defines the flow value regarding its various states and general transition behaviors such as initialization, copy, and meet.
Then,
the flow functions are designed to describe how specific IR instructions can transfer the state of the flow value.
To improve the efficiency of analysis,
Sharlit can receive a set of simplifying rules that merges several flow functions along a path into a single function.
Such a design idea has been integrated into the dataflow engine of SOOT~\cite{Vall1999Soot}, a Java analysis framework.

\textit{\textbf{Taint Analysis.}}
Consider task V in Tab.~\ref{tab:analysis-tasks}.
Taint analysis is to determine how taint data can influence the program through both data flow and control flow.
In order to fluently customize various taint analysis problems,
a query language \textit{fluentTQL} is presented~\cite{Piskachev2022TQL}.
Firstly, it enables a developer to establish the relationship between functions and the three fundamental elements of taint analysis: source, sanitizer, and sink.
Furthermore, it allows for the description of the behavior of ``propagators,'' which are functions that implicitly propagate the taint data.
Additionally, a specific \textsf{and()} operator is provided to allow the combination of two related taint flows to build a more intricate taint problem.

\textit{\textbf{Value-flow Analysis.}}
Value flow analysis determines if a source value may flow to a sink site.
A query language called VFQL~\cite{VFQL} is introduced to address the various source-sink problems.
To begin with,
VFQL allows the user to group the IR nodes into labels, representing different ways of producing and using value.
For instance, it categorizes values into four primitive labels (\textsf{const}, \textsf{copy}, \textsf{operator}, and \textsf{join}) depending on the four ways they are formed.
Subsequently, a simple VFQL query may be generated by specifying the labels of the source and sink. 
This query can be enhanced with a \textsf{where} clause to define further whether or not a label should appear along the trace.
In addition, it differentiates between the "exists" and "for-all" queries of traces using two quantifiers, namely \textsf{to} and \textsf{all-to}.

\textit{\textbf{Instrumentation.}}
PQL'05~\cite{Martin2005PQL} provides the capability to use static analysis to assist instrumentation,
thus ensuring code safety during runtime.
When a potential vulnerable pattern is matched, 
the handling code specified by users could be inserted before the vulnerable statement.

\textit{\textbf{Query-based Analysis.}}
Based on our ongoing discussion,
all of the query designs assume an underlying IR.
Nevertheless, in the early stages, scholars have proposed that static analysis can be conducted through relational queries regarding program relations~\cite{Linton1984Relational, Jarzabek1998PQL}.
In particular, program relations can be extracted to a relational DB by preliminary analysis,
with no assumptions for the underlying IR.
Afterwards, 
the static analysis algorithm is defined in a declarative manner, 
and this will subsequently be transformed into queries on the database. 
As an early attempt,
CodeQuest~\cite{Hajiyev2005CodeQuest} uses Datalog to declare program queries and compiles them into SQL.
Afterward, 
CodeQL~\cite{avgustinov2016QL} is a more powerful Datalog-based static analysis tool.
It formulates program relations as two types of predicates.
The Extensional Database (EDB) is a mapping of the program relations already stored in the database, such as basic def-use information and type information.
The Intensional Database (IDB) describes the high-level program relations derived from the existing ones, such as pointer and dataflow information.
an IDB predicate can be defined in an object-oriented way through pre-defined or user-defined queries, with the form of ``select-where-from''.
Meanwhile, other works also develop their analyzer using a graph database.
For example, GRUPRO~\cite{Ebert2002GUPRO} utilizes GreQL as its query language, while Wiggle~\cite{Urma2015Wiggle} employs Cypher.
These approaches provide high expressiveness and flexibility,
however,
they also encompass certain drawbacks.
Without assuming the existence of an underlying IR, 
it becomes challenging to conceptually understand various program relationships. 
This places an additional learning burden on developers.
In addition, the declarative programming model poses difficulties in optimizing the performance of particular analysis algorithms.

\section{IR Preprocessing}\label{sec:preprocess}

\begin{figure}[t]
  \centering
  \includegraphics[width=0.48\textwidth]{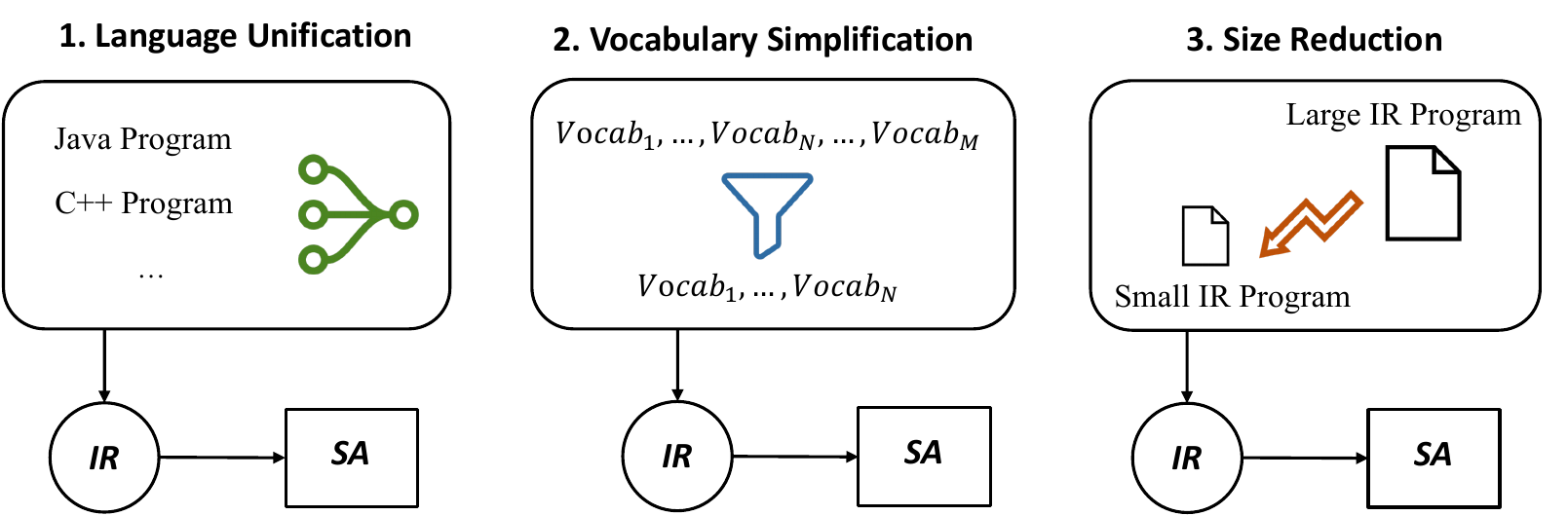}
  \caption{An overview of IR preprocessing techniques}
  \label{fig:preprocess}
\end{figure}

\begin{figure*}[t]
  \centering
  \includegraphics[width=0.7\textwidth]{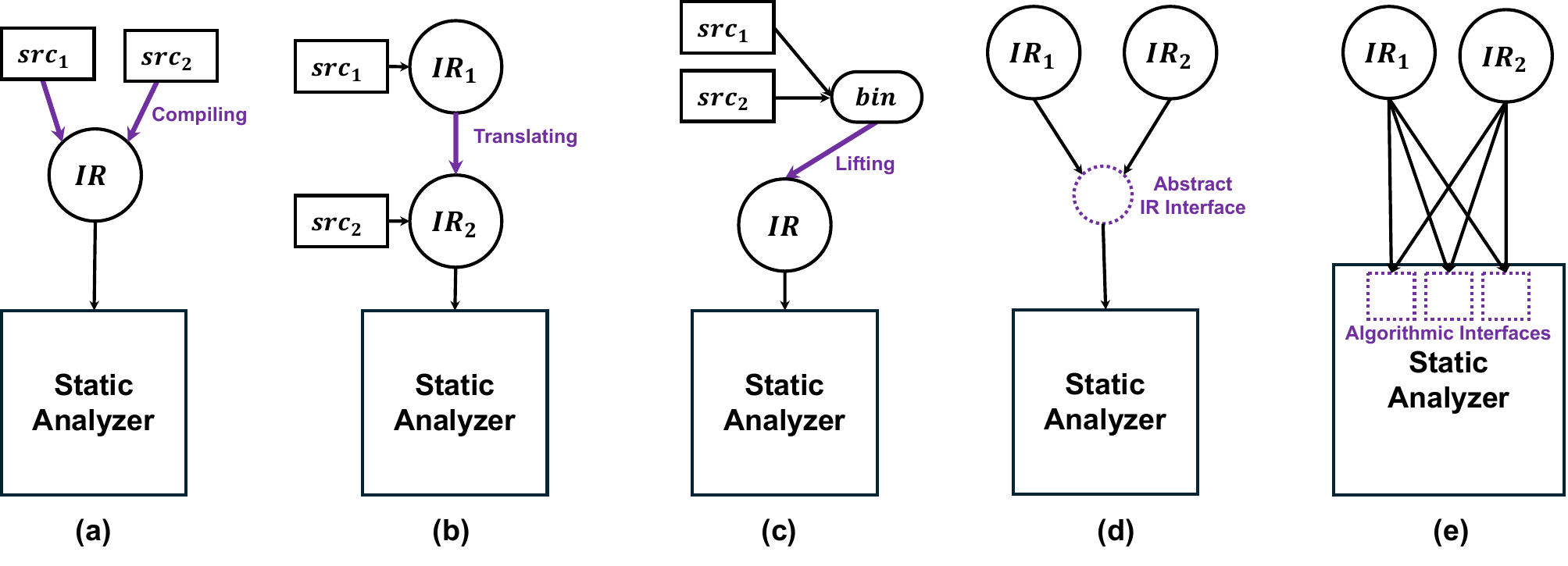}
  \caption{Illustration of different IR unification techniques}
  \label{fig:union}
\end{figure*}

To analyze a program,
the static analyzer needs to take the corresponding IR program as input.
This section showcases the advantages of performing preprocessing on the IR programs before it is fed into the static analyzer.
As depicted in Figure~\ref{fig:preprocess}, 
we categorize the preprocessing techniques into three sub-categories, 
according to distinct aspects of their advantage to static analysis.
In what follows,
we briefly introduce these three sub-categories.
\begin{itemize}
\item 
Firstly,
a static analysis framework is inherently tied to a particular IR.
Hence, the \textbf{\textit{unification}} technique is designed to enable static analysis to treat various languages in a unified manner,
using IR as a bridge (Fig.~\ref{fig:preprocess} (a)).
\item
Secondly,
a modern IR could encompass a high level of complexity, which places a mind burden on developers.
The complexity arises from both syntax (\S~\ref{sec:syntax}) and vocabulary (\S~\ref{sec:voc}).
For instance, 
syntax can be instantiated to complicated graph structures,
while vocabulary can be reflected in the large instruction set of IR.
Therefore,
the technique \textbf{\textit{simplification}} is to reduce the complexity of IR so that the subsequent static analysis can focus on a simpler IR,
such as an IR with a simpler instruction set (Fig.~\ref{fig:preprocess}(b)).

\item
Thirdly,
as modern software grows huge, 
the size of their corresponding IR could be respectively large.
In this context,
a way to improve the speed of static analysis is to perform \textbf{\textit{reduction}} on IR,
which turns a large IR into a smaller one (Fig.~\ref{fig:preprocess}(c)).
\end{itemize}

The detailed approaches to each technique will be elaborated in the next sections.

\subsection{Unification}\label{subsec:unify}

In order to achieve unification,
researchers have employed different approaches,
including compiling, translating, lifting, and using interfaces.
We first introduce each approach and then summarize the scenarios of choosing different approaches.

\textbf{\textit{Approach 1: Compiling.}}
The most prevailing approach is to utilize the infrastructure of compilers.
As shown in Fig.~\ref{fig:union}(a),
as long as a language can be compiled into an IR,
the static analysis developed on this IR can be enabled to support this language.
For instance,
recall Tab.~\ref{tab:voc},
it can be observed that a popular compiler IR such as LLVM IR can often support multiple languages.

However,
a static analyzer is typically developed for the IR compiled from a particular language,
such as LLVM IR generated from C++. 
Besides,
variations exist in the IRs produced from different languages, 
necessitating the need for adaptation efforts.
Garzella et al.~\cite{Garzella2020VMCAI} share such experiences in adapting SMACK, 
a C static analyzer based on LLVM IR,
to other languages within the ecosystem of LLVM.
To adapt the IR captured from a new language,
the authors point out two core aspects to consider.

First, new IR constructs.
IR, as a highly capable programmable system,
employs only a subset of its constructs when a specific language is lowered to it.
We denote this subset used by a language $L$ as $C_{L}$.
Hence, when expanding a static analyzer from $L_1$ to encompass $L_2$, 
it becomes essential to handle the additional constructs in $C_{L_2} - C_{L_1}$. 
This is demonstrated through various examples associated with the Rust compiler. 
In comparison to Clang, 
the Rust compiler generates a greater number of structure operations, 
employs LLVM's intrinsics for secure arithmetic operations, 
and optimizes storage by compressing structures into integers.
These new structures are managed either by reducing them to existing structures or by devising new analysis logic tailored to them.
The adaptation process for the static analyzer is not ad hoc; 
in fact, 
it gradually enables the analyzer to become complete for several languages and eventually achieve completeness for the entire IR.

Second,
un-modeled libraries.
Languages are typically accompanied by extensive standard libraries and runtime libraries. 
For instance, the specific implementations of a "for-all" loop may vary across different languages. 
Analyzing these libraries can be approached by compiling their implementation code into IR and linking it with other components. 
However, 
the complexity introduced by library function implementations, 
driven by performance optimizations like assembly code or specialized techniques,
poses a challenge. 
Consequently, 
the direct approach's resulting IR is not amenable to effective static analysis.
Therefore, 
alternative approaches only preserve the core semantics of the libraries with some human-written specifications.
For instance, 
one approach is to design a stub, which is a simplistic implementation, 
for these library functions. 
Another approach is to design a summary for a specific analysis, 
where a summary provides a concise representation of the behavior and effects of a library function.

\textbf{\textit{Approach 2: Translating. }}
As shown in Fig~\ref{fig:union}(b),
the second approach~\cite{Zhang2006ISSTA, Bartel2012Dexpler, Arzt2016SOAP, CIL2Bytecode} involves translating other IRs into the IR used by the static analyzer.
Zhang et al.~\cite{Zhang2006ISSTA} were the first to investigate this concept, 
which they named \textit{Analysis-Preserving Language Translation (APLT)}.
They contend that the process of \textit{Compilation-Preserving Language Translation} is exceedingly challenging and unnecessary for static analysis, 
which primarily aims to identify error states associated with certain program paths.  
Thus, APLT prioritizes maintaining the control- and data-flow of the program over strictly adhering to semantic equivalence. 
However, it does allow for minor deviations from the original program semantics in certain respects.
For evaluation,
the authors implemented a C-to-Java translator to reuse a security static analyzer on Java IR.
Specifically, 
the C-style function pointers are translated into multiple overloads of a method in the same class to ensure that the program's control flow remains unchanged. 
Additionally,
Java features, such as \textsf{setjmp/longjmp}, inline assembly, and arbitrary pointer arithmetic, are compromised during the translation process.
To validate the correctness of the translation,
the authors employ another counterpart analysis implemented on C to differentially test the analysis results. 

Dexpler~\cite{Bartel2012Dexpler} performs the translation of Dalvik bytecode (DEX) into Jimple,
enabling Soot to analyze Android projects.
Out of the total 237 instructions, 
the authors identified 20 instructions solely used for optimization and code generation, 
which do not appear in the application code and therefore require no attention. 
The translation successfully covers the rest five major instruction categories: 
move, branch, getter/setter, method invocation, and arithmetic operations.
In order to validate the correctness of the translation, the authors employ Soot's capability to convert Jimple back to Java bytecode. They then compare the results of executing test cases produced using Javac directly to those changed using Dexpler and Soot. 
Subsequently, they meticulously analyze the call graph generated by Soot, further increasing the confidence.

Furthermore,
in order to extend the capability of Soot for C\#,
Arzt et al.\cite{Arzt2016SOAP} introduced a CIL~\cite{ECMA335} to Jimple translator.
To address the distinctive features of C\#,
the authors investigate the corresponding representations in CIL and choose appropriate translation strategies.
\begin{itemize}
\item 
On the one hand,
it is discovered that certain features are already transformed by the compiler,
eliminating the necessity for special treatment.
For example,
C\# operator overloading is implemented as regular function calls,
and class properties like \textsf{get} and \textsf{set} are converted into normal member functions like \textsf{getter} and \textsf{setter},
requiring no modification to the original analysis engine.
\item
On the other hand,
when the compiler's implementation of a feature incorporates specific constructs in the IR,
the developers would take note of these intricacies.
For example, C\# delegates allows functions to be passed as parameters and stored in objects,
which helps the use of callback mechanisms.
In CIL,
delegates are implemented as a class that consists of a function pointer,
a constructor accepting a function pointer,
and a builtin function called \textsf{Invoke}.
To create a delegate, a function pointer is loaded and passed to the constructor to start the delegate object. To use a delegate, the stored function pointer is called through \textsf{Invoke}.
Therefore,
the intricacies involves managing function pointers, 
as discussed in \cite{Zhang2006ISSTA},
and explicitly defining the \textsf{Invoke} function in Jimple.
\end{itemize}

\textbf{\textit{Approach 3: Lifting.}}
It is known that binary lifting elevates binary programs such as x86 form to higher-level representations for better understanding.
Specifically,
these outcomes include customized~\cite{Egalito} or existing~\cite{SecondWrite, Inception, BinRec} IRs.
It would be ideal if these lifted IRs could be used out-of-the-box for static analysis tasks (Fig.~\ref{fig:union}(c)),
as obtaining the binary of a software project is easily achievable,
while intervening in the compilation system or implementing IR translation would be more cumbersome.
However, such a technical approach poses challenges due to the loss of information during the compilation and code generation process.
A recent study~\cite{Liu2022Sok} demonstrates that four modern lifters~\cite{BinRec, RetDec, mctoll, McSema} that generate LLVM IR performed poorly in the task of pointer analysis,
producing almost meaningless results.
The main gap lies in recovering type information for variables and functions from stripped binaries, especially pointer types.
The study further shows that the pointer analysis results can be significantly improved by recovering or directly enabling debug information in binary.
Another recent work proposed a new lifter, Plankton~\cite{Plankton} to handle binaries with debug information and utilize the produced IR for value flow bug detection,
achieving only a 17.2\% loss in precision compared to directly compiled IR.
It introduces two algorithms that utilize debug information.
The first one recovers source-level variables by separating the stack memory, while the second one recovers type information by applying a fixed point algorithm to eliminate type conversions and maximize the type correctness level.
Overall, 
static analysis using lifted IR remains an open problem with many challenges awaiting future research.

\textbf{\textit{Approach 4: Using Interfaces. }}
Different from the previous three approaches,
the last group of works~\cite{Devanbu1992GENOA, Devanbu1994Aria, Hayes2000StarTool, Strout2005IRIndep, Zhao2016Ontology, Teixeira2021SOAP}
attempt to remove the constraint of associating static analysis with a particular IR.
An adaptation interface is proposed to bridge the gap between static analysis and multiple IR,
following the concept of generic programming.
Consequently, 
the design principles for the interface can be divided into two categories:
IR-centered and Client-centered.

\noindent\textit{\textbf{IR-centered Interface.}}
As shown in Fig.~\ref{fig:union}(d),
this approach proposes an abstract IR,
which is independent of the details of the specific IRs.
Subsequently,
the static analysis algorithm is implemented on generic interfaces of this abstract IR.
GENOA~\cite{Devanbu1992GENOA} represents the initial endeavor.
It introduces an abstract IR called \textit{g-tree}, which encapsulates the shared language concepts of the C language family.
Then,
the different analyses are described with the interfaces of g-tree,
which is IR neutral.
When adopting a new IR,
one needs to implement the interfaces using the APIs specific to this IR.
For example,
to implement the abstract global variable in g-tree,
the process involves implementing corresponding interfaces to retrieve attributes of global variables in a specific IR,
such as name, type, and accessibility.
GENOA was assessed in several syntactic-level code search tasks, 
and subsequent studies have extended this method to encompass more static analysis scenarios.

Aria~\cite{Devanbu1994Aria} selects Reprise~\cite{Rosenblum1991Reprise},
another abstract IR similar to the aforementioned g-tree.
Based on the interfaces of Reprise to identify the jump targets associated with the control flow statements,
Aria implements a generic algorithm to construct a control dependence graph (CDG) for both C and C++.

PATO~\cite{Zhao2016Ontology} suggests utilizing knowledge graphs as the abstract IR,
to depict program concepts and their interconnections.
The subsequent static analysis then becomes declarative, 
much like CodeQL~\cite{avgustinov2016QL}.
Some simple analysis algorithms, 
such as Anderson-style pointer analysis and canonical loop analysis,
can be implemented within the PATO system.
However, 
PATO does not demonstrate the ability of a knowledge graph to abstract multiple languages into a unified interface. In fact, 
in order to support C analysis, 
PATO extracted approximately 250 concepts from the C99 specification to create the  C knowledge base. 
It can be inferred that if another language were to be introduced, 
the capability for the 2-language generic analysis would depend on the intersection of concepts between the 2-language knowledge base. 
The more languages introduced, 
the fewer generic analyses can be supported.

The LARA project~\cite{Teixeira2021SOAP} emphasizes the abstract IR should be designed from the perspective of language paradigms, 
rather than characteristics of specific languages. 
In particular, 
the authors proposed a language specification for the OO paradigm to establish an abstract OO IR for languages including C, C++, Java, and JavaScript.
Next,
the authors successfully implemented various OO metrics~\cite{CK1994metrics, LH1993metrics}
 to evaluate the cohesion, coupling, and maintainability of OO programs.

In summary, 
the overall principle of IR-centered interfaces is to abstract common concepts between languages as much as possible, 
without assuming downstream static analysis clients. 
However, 
the reality is that there are significant semantic differences between IRs that cannot be unified in this way. 
At the same time, 
all of the cases mentioned above show that the abstract IRs they propose are essentially an abstraction layer for specific language features,
and thus can only serve a portion of downstream analyses.

\noindent\textit{\textbf{Analysis-centered Interface.}}
As shown in Fig.~\ref{fig:union}(e),
this approach decomposes an analysis algorithm into several generic interfaces.
Subsequently,
such interfaces can be implemented differently using the APIs of specific IRs.

This idea was first investigated in the development of StarTool~\cite{Hayes2000StarTool},
a code refactoring tool aiming to support multiple languages.
During its evolvement,
its design principle shifted from an IR-centered approach to an analysis-centered approach.
There are two reasons for this.
Firstly, extracting the abstract IR interfaces from each specific IRs without considering the client analysis would introduce unnecessary adaptations.
This is because a specific analysis only touches a limited set of program properties.
Secondly, an IR-centered interface may not fully exploit the capabilities of an IR system. 
For instance, to implement the functionality of filtering particular types of IR instructions, 
the IR-centered approach necessitates exporting interfaces that provide the type of each IR instruction.
Nevertheless, IRs that offer direct filter functionalities are unable to fully harness their potential.
Therefore,
the analysis-centered interface intends to summarize the minimum prerequisites of a static analysis on its IR,
allowing various IRs to satisfy these requirements in a flexible manner. 
Moreover,
the authors highlight that a complicated analysis can be achieved by reusing interfaces designed to satisfy existing analyses.
 
OpenAnalysis~\cite{Strout2005IRIndep} extends this idea to several practical alias analyses and data flow analyses.
Alias analysis reasons about which memory location a memory reference expression may or must access.
To make alias analysis generic,
the authors first propose interfaces regarding different types of \textit{memory locations}.
For example,
statically allocated locations abstract local and global variables,
dynamically allocated locations abstract heap and stack objects,
subset locations abstract array elements and structure fields,
etc.
Next,
the interfaces of \textit{memory reference expressions} are proposed to abstract the memory effect of a statement.
Specifically,
given a specific IR statement,
several expressions would be emitted regarding the reference, dereference, and address computation of the mentioned memory locations recognized in the statement.
Finally,
the downstream alias analysis is developed by performing several generic iterations on alias-related statements and traversing the expression interfaces.
To handle reaching constant analysis,
the memory location interfaces are reused,
and the memory reference expression interfaces are just replaced with the constant value interfaces regarding the definition, usage, and computation of (non-) constant values.

\begin{table}[]
\renewcommand\arraystretch{1.4}
\centering
\begin{tabular}{c|c|c}
\hline
\textbf{Approach} & \textbf{Advantages}                                      & \textbf{Disadvantages}      \\ \hline
Compiling         & Convenience                                              & Rely on ecosystem \\ \hline
Translating       & \multicolumn{1}{c|}{\multirow{2}{*}{Bridge ecosystems}} & Validation requirement      \\ \cline{1-1} \cline{3-3} 
Lifting           & \multicolumn{1}{c|}{}                                    & Information loss            \\ \hline
Using interfaces         & Flexibility                                              & Generality loss             \\ \hline
\end{tabular}
\caption{Comparison between different unification approaches}
\label{tab:unify-approaches}
\end{table}

\textbf{Comparison between approaches. }
In Tab.~\ref{tab:unify-approaches},
we list the keywords regarding the pros and cons of the four unification approaches.
The compiling strategy utilizes the current compiler infrastructure, 
but is limited to accessing just the IRs within that ecosystem.
Meanwhile, 
the translation and lifting methods can connect diverse ecosystems, 
but one must address the associated concerns of correctness and loss of information.
Ultimately, employing generic interfaces enables the flexible utilization of the current IR systems without altering them. 
However, this approach reduces the expressiveness of the underlying IR,
which is detrimental to a framework aiming to provide generic static analysis capabilities.

\subsection{Simplification}\label{subsec:simplify}
IR simplification lowers the complexity of IR itself to reduce the burden of development.
According to an empirical study on 115  static analysis developers~\cite{Do2020TSEDebugSA},
the most prevailing error when debugging a static analyzer is to handle the corner cases regarding the complex IR constructs.
Such complexity arises from the syntax and vocabulary employed in the IR. 
In terms of vocabulary, the IR typically encompasses a wide array of instructions (operations). 
Additionally, there are numerous library functions that are referenced as function calls without explicit definitions. 
The intricacy of the syntax can be attributed to the combination of the aforementioned vocabularies, 
as well as the intricate graph structures, 
such as CFG structures.

\begin{table}[t]
\renewcommand\arraystretch{1.4}
\begin{tabular}{c|l}
\hline
\textbf{Granularity}                                                        & \textbf{Optimizations}                                                                                                                                \\ \hline
Single Instruction                                                     & \textsf{Lower-Switch}, \textsf{lower-Invoke}, \textsf{Lower-Atomic}                                                                                                     \\ \hline
\begin{tabular}[c]{@{}l@{}}Instruction Sequence\end{tabular} & \textsf{Inst-Combine}                                                                                                                          \\ \hline
Graph                                                             & \begin{tabular}[c]{@{}l@{}}\textsf{Simplify-CFG}, \textsf{Merge-Return}, \textsf{Code-sinking} \\ \textsf{Loop-Simplify}, \textsf{Loop-Rotate}, \textsf{Loop-Reduce}\end{tabular} \\ \hline
\end{tabular}
\caption{Examples of LLVM optimizations that can be utilized for simplification}
\label{tab:opt}
\end{table}

\textbf{\textit{Compiler Optimization.}}
Fortunately,
researchers have realized that many compiler optimizations share the same goal as simplification.
As shown in Tab.~\ref{tab:opt},
we have collected some representative optimization passes from the famous LLVM compiler.
They simplify the IR at different levels of granularity.
Firstly,
a single complex instruction could be optimized for simpler instructions.
For example,
\textsf{Lower-Switch} optimizes the switch instruction into several equivalent branch instructions.
Secondly,
\textsf{Inst-Combine} (a.k.a. peephole optimization) optimizes a sequence of consecutive instructions into fewer and simpler instructions, 
often involving algebraic simplifications. 
With proper utilization, 
it can enable static analyzers to make fewer assumptions about the computational expressions in the IR.
Thirdly,
many optimizations are utilized to simplify the assumption of the graph structure.
For instance,
after the \textsf{Merge-Return} optimization,
the CFG can be ensured to have only one return,
which could ease the implementation of dataflow analysis~\cite{Zhao2018PSEG}.
Similarly,
\textsf{Loop-Simplify} simplifies loop structures to a canonical form~\footnote{https://llvm.org/docs/LoopTerminology.html}.

In practice, 
different opinions are held on the usage of compiler optimizations.
Specifically, 
SVF~\cite{Sui2016SVF} recommends using unoptimized IR, 
but it strives to support as many structures of IR as possible.
PhASAR~\cite{Schubert2019Phasar} suggests an approach to debug static analyzers in the context of various optimizations:
Initially,
thoroughly debug the static analyzer on a set of test cases without any optimizations.
Subsequently,
the optimizations are activated, 
and differential testing is performed, 
with the results from unoptimized IR serving as the oracle for comparison.
On the contrary,
there are several works that employ optimizations.
For instance,
PINPOINT~\cite{Shi2018Pinpoint} employs the three lines of optimizations in Table~\ref{tab:opt} to reduce the IR structures that necessitate handling. 
Additionally, 
it employs loop unrolling~\cite{Dongarra1979unroll} to control the number of paths involved in practical path-sensitive bug detection.
SeaHorn~\cite{Arie2015SeaHorn} reports using \textsf{Inst-Combine} and \textsf{Simplify-CFG},
In addition,
it incorporates a step that converts aggregate operations into scalar operations, aligning with the first level of granularity.
SMACK~\cite{Montgomery2016SMACK} also makes use of LLVM optimizations,
with the purpose of enhancing the performance and accuracy of the following verification process.

However, 
the impact of compiler optimization on static analysis remains an open question and has not been systematically addressed.
Namjoshi et al.~\cite{Namjoshi2018SAS} first presents a framework to formally analyze the impact of compiler optimizations on the precision of abstract interpretation-based static analyses~\cite{Cousot1977AI}. 
Given a semantics-preserving program optimization that transforms a source program \textsf{S} to a target program T, 
their approach requires specifying a bisimulation relation \textsf{B} that establishes the semantic equivalence between \textsf{S} and \textsf{T}. 
At each step of the static analysis on \textsf{S}, 
the current abstract state $a_\textsf{S}$ is transferred to the corresponding abstract state $a_\textsf{T}$ in \textsf{T} by applying the bisimulation relation \textsf{B}. 
The analysis is then performed on $a_\textsf{T}$ in \textsf{T}, 
yielding a new abstract state $a'_\textsf{T}$. 
This new state is then reversely transferred to a state $a'_\textsf{S}$ in \textsf{S}  by applying the inverse bisimulation relation $\textsf{B}^{-1}$. 
By comparing $a'_\textsf{S}$ with $a_\textsf{S}$ which is obtained by directly running the analysis on $S$, 
the impact of the optimization on analysis precision can be evaluated.

\textbf{\textit{Lifting Inline Assembly. }} 
Inline assembly is commonly used to achieve low-level functionality like multicore semantics, performance optimization, and hardware control~\cite{Rigger2018Inline}.
It is typically encoded in the IR as a raw string of hardware assembly, which poses a challenge for static analysis~\cite{Xie2007Saturn}.
To uncover hidden semantic information within inline assembly, TInA~\cite{Recoules2020TInA} employs binary lifters to elevate inline assembly into analyzable IR, enabling direct analysis by existing static analyzers. 
Its follow-up work RUSTInA~\cite{Recoules2021RUSTInA} discovers numerous bugs where inline assembly did not properly conform to expected interfaces.

\textbf{\textit{Mocking Libraries.}}
Lastly,
to tackle the complexity of libraries,
researchers have proposed mocking their behaviors.
This approach often starts with a lightweight analysis to extract the behavior of the library. 
Some ``stubs'' are then generated in the IR to simulate the library's essential behavior.
Consequently,
the subsequent static analysis is able to conduct correct reasoning about this library.
A typical scenario is the analysis of web frameworks,
which encapsulates common development tasks such as request routing, form handling, authentication, logging, and caching.
Their presence can obscure data flow information and call edges, 
making it difficult for static analysis tools to fully understand the code.
Existing efforts~\cite{Tripp2009TAJ, Wang2020ANTaint, Antoniadis2020JackEE, Chen2023Jasmine} extensively utilized the mocking approach to address such complexity.

To address the issue of multiple forms corresponding to the parameters of controller methods in the Struts framework, 
TAJ~\cite{Tripp2009TAJ} analyzed the cast operations in the controller code to selectively extract type-compatible form data. 
For each matched form data type, 
TAJ simulated the creation of its object and passed it to the controller method,
thereby achieving an over-approximation effect for taint analysis.
Following the designation of the controller method as the taint analysis entry point, 
ANTaint~\cite{Wang2020ANTaint} inserts additional statements that serve to label the incoming and outgoing values of the controller method with respect to frontend requests, external invocations, and database interactions as taint sources and sinks, respectively.

JackEE~\cite{Antoniadis2020JackEE} establishes a vocabulary for the common concepts of Java web frameworks,
such as Servlet, Controller, and Bean.
Then,
it builds Datalog rules to identify the concepts from the program and infer useful information for static analysis,
mainly concerning the entry point methods and bean dependency.
Consequently,
JackEE generates the mocking code to supplement the necessary information for the original IR.
Similar to TAJ~\cite{Tripp2009TAJ},
it mocks the invocation of the entry point by creating the associated object and type-compatible arguments.
For Bean objects,
JackEE identifies dependency injections from Annotations and configuration files,
explicitly performing them in the IR for subsequent static analyzer recognition of the connectivity between Bean objects.
A subsequent work,
Jasmine~\cite{Chen2023Jasmine},
proposes a \textit{weave process} to mock the intricate invocation process behind target service methods,
by utilizing the Annotations of SpringAOP such as \textsf{@Around}, \textsf{@Before}, and \textsf{@After}.


\subsection{Reduction}\label{subsec:reduction}
By reducing the large IR programs into small ones,
the reduction technique is vital in accelerating static analysis.

First,
compiler optimization plays a crucial role in accomplishing this objective by eliminating "dead" IR elements and structures.
They burden static analyzers with unnecessary workloads, 
thereby diminishing their efficiency. 
Typically,
the optimization \textsf{Dead-Code-Elimination} sweeps out unused instructions from the IR.
Similarly,
other IR elements such as types, arguments, and globals could also be removed by corresponding optimization passes.
At the CFG level,
passes like \textsf{Simplify-CFG} and \textsf{Loop-Deletion} remove the unreachable blocks and loops with no impact on program execution.

Furthermore,
researchers customize specific reduction techniques for particular static analyses, 
aiming to ``indirectly'' reduce the size of IR.
Static analyzers, 
such as abstract interpreters~\cite{Gange2016Crab}, 
symbolic executors~\cite{Cadar2008Klee}, 
and path-sensitive bug detectors~\cite{Xie2007Saturn,Shi2018Pinpoint}, 
explore program paths, 
and in certain instances,
employ constraint solvers~\cite{Z3} to ascertain the satisfiability of path conditions.
To tackle the scalability issue arising from path explosion,
TRIMMER~\cite{Ferles2017Trimming} proposes inserting necessary unsafety conditions into the IR to preemptively terminate the execution of these static analyzers.
It formalizes the problem on a C-style language with \textsf{assume} and \textsf{assert} operations.
Then for a program $p_1....p_n$, 
a backward analysis is performed to collect each weakest precondition $\Phi_{i}$ for $p_i...p_n$ to safely terminates without violating the assumptions or assertions.
Subsequently,
a statement \textsf{assume $\neg \Phi_{i}$} is instrumented before the sensitive statement $p_i$ (e.g. a loop and a branch) to filter out unnecessary path explorations.
By observing that the control flow directly terminates when the assertion fails,
Akash et al.~\cite{Akash2014FMCAD} propose exposing the assertions to the \textsf{main} function by both partially inlining the functions and unrolling the loops,
thus avoiding redundant computations on loop invariants or function summaries.

Another way to conquer the large IR size is to divide the IR into parallelizable partitions.
Christakis et al.~\cite{Christakis2022InputSplit} motivate that it can horizontally enhance the efficiency of static analyzers without modifying the individual tools.
They study two module-level partitioning strategies.
One naively divides the modules into disjoint partitions,
and the other utilizes dependency information to obtain the partitions that can overlap.
By evaluating the two strategies using three static analyzers on multiple benchmarks,
it's found that for linter-like and intra-procedural analyses, 
the naive strategy could already preserve accuracy and even achieve better efficiency.
However,
for more complex analyses,
the dependency-based strategy is necessary to reduce accuracy loss.
In future work, 
a more fine-grained partition strategy needs to be proposed to achieve better parallelism. 
At the same time, 
balancing the loss that controls accuracy is an important metric to consider.

\section{Discussions}\label{sec:discussion}
This is the first systematic review of Intermediate Representations(IR) for static analysis.
We conclude by summarizing its significance for different individuals within the static analysis community.

\subsection{Suggestions for Learners}
Static analysis learners can now comprehend how the IR serves as the programming language for static analysis.
In addition,
an individual can get a comprehensive understanding of a static analysis framework by first grasping its IR.
Specifically,
one can first identify what kind of syntax and vocabulary this IR possesses.
Then, 
it is beneficial to learn the IR queries provided by the framework to make implementation of analyses easier.
Finally,
by understanding the IR transformations involved in the framework,
one can be more aware of the potential assumptions of this IR.

In the process of studying static analysis, readers can further delve into its algorithms~\cite{Mller2011StaticPA, CSURParametric, CSURHeap}, applications~\cite{IBMSASec, Sufatrio2015Android}, or practical aspects~\cite{Nachtigall2022Metrics, CSURAlarm}.

\subsection{Takeaways for Practitioners}
For practitioners of static analysis,
we have summarized the core spirit of this survey in Tab.~\ref{tab:summary},
highlighting how to leverage the IR to enhance the three overarching objectives of static analysis from various viewpoints.

For versatility, we design a suitable IR vocabulary based on the main target language of static analysis (\S~\ref{sec:voc}). Next, through the IR unification technique (\S~\ref{subsec:unify}), we can make the static analysis system encompass more languages.

For performance, we design an appropriate syntax to highlight program relationships, enabling the analysis to be conducted effectively and efficiently(\S~\ref{sec:syntax}). Then, we can utilize reduction techniques (\S~\ref{subsec:reduction}) to reduce the size of the IR, thereby further accelerating the analysis process.

For productivity, We can encapsulate IR manipulation or static analysis algorithms into query interfaces(\S~\ref{sec:query}). Furthermore, we can employ simplification techniques(\S~\ref{subsec:simplify}) to reduce the complexity of IR to reduce the workload of static analysis.

\begin{table}[t]
\renewcommand\arraystretch{1.2}
\centering
\begin{tabular}{c|ccc}
\hline
\multirow{2}{*}{\textbf{\textit{Category}}} & \multicolumn{3}{c}{\textbf{\textit{Benefit to Static Analysis}}} \\
\cline{2-4}
& \multicolumn{1}{c|}{\textbf{Versatility}} & \multicolumn{1}{c|}{\textbf{Performance}} & \textbf{Productivity} \\
\hline
\textbf{IR Syntax} & \multicolumn{1}{c|}{} & \multicolumn{1}{c|}{\checkmark} & \\
\hline
\textbf{IR Vocabulary} & \multicolumn{1}{c|}{\checkmark} & \multicolumn{1}{c|}{} & \\
\hline
\textbf{IR Query} & \multicolumn{1}{c|}{} & \multicolumn{1}{c|}{} & \checkmark \\
\hline
\textbf{IR Preprocessing} & \multicolumn{1}{c|}{\checkmark} & \multicolumn{1}{c|}{\checkmark} & \checkmark \\
\hline
\end{tabular}
\caption{A summary on how IR enhances static analysis}
\label{tab:summary}
\end{table}

\subsection{Future Directions for Researchers}
IR serves as the cornerstone of static analysis techniques and provides benefits in various aspects.
We summarize some future research directions for static analysis and IR.

\textbf{Direction 1. Analyzing more domain-specific languages.}
IR has been designed across various domains such as machine learning, parallel computing, and hardware design, 
while IR frameworks like MLIR allow for multiple IR abstraction levels (\S~\ref{subsec:domain}).
One challenge is reusing old static analysis types in new contexts, 
while another is dealing with unique analysis types in new scenarios,
which may require the coordination of IRs at various abstraction levels.
A new static analysis framework is needed for these two challenges. 
For the first challenge,
researchers can draw inspiration from the ideas of generic static analysis in \S~\ref{subsec:unify}
For the second challenge,
research may further extend the idea of SAIL~\cite{dillig2009sail}.

\textbf{Direction 2. More complete and sound static analysis.}
After the IR is settled, 
an IR-based static analyzer is like handling multiple ``switch cases'' related to vocabulary (\S~\ref{sec:voc}).
It is crucial to ensure that such handling is complete and sound,
i.e. more cases being handled and each case being handled correctly.
To achieve this goal,
there are two possible directions to explore.
The first direction is to generate a large number of IR programs and construct oracles related to static analysis. 
However, 
current program generation techniques~\cite{Yang2011Csmith, YARPGEN} primarily start from source code and still lack diversity in IR. 
Moreover, compared to oracles used in compilers, such as I/O behaviors~\cite{EMI}, 
designing oracles for static analyzers is more challenging.
The second direction is extracting IR semantics underlying the syntax and vocabulary of IR,
and then use it as a ground truth to validate static analyzers.
Existing works have already defined IR semantics for popular IRs in different approaches.
e.g.,
KLLVM~\cite{Li2020KLLVM} under the $\mathbb{K}$ framework~\cite{Rou2017AS},
and Vellvm~\cite{Zhao2012Vellvm}, Crellvm~\cite{Kang2018CreLLVM}, and VIR~\cite{Zakowski2021VIR} established under Coq.
Additionally,
IR semantics has been successfully adopted in validating compiler optimizations~\cite{Lopes2015ProvablyCP, Lopes2021Alive2BT, Bang2022SMTBasedTV},
which is achieved via checking IR equivalence before and after the optimization.
However,
such a mechanism cannot be applied to static analyzers.
Taneja et al.~\cite{Taneja2020TestingSA} performs symbolic abstractions of analysis results based on IR semantics,
which can be used to validate the results of existing static analyzers.
However, this approach only supports basic data flow analyses.

\textbf{Direction 3. Synthesizing static analyzers.}
Both IR-centered and Analysis-centered query designs have significantly simplified the burden of static analysis development (\S~\ref{sec:query}).
Program synthesis is a promising direction toward a more automated static analyzer.
Existing works have extensively explored the synthesis of Datalog-based static analyzers~\cite{Si2018FSE, Raghothaman2019POPL, Nai2021Sporq},
whose difficulty features the combination of program relations.
However, 
even with the query interfaces,
synthesizing an IR-based static analyzer can be considered a more challenging direction,
similar to an algorithmic synthesis problem~\cite{Sun2023OOPSLA, Farzan2017PLDI}.

\textbf{Direction 4. Evaluating the effect of IR preprocessing.}
In the IR preprocessing techniques we discussed, the IR often undergoes transformations before being used as input for static analysis (\S~\ref{sec:preprocess}).
Can the benefits of these transformations be measured quantitatively, 
and are there any disadvantages associated with them?
While previous research has explored binary similarity analysis~\cite{Ren2021BinTuner}, static analysis has not been thoroughly investigated.

\textbf{Direction 5. IR post-precessing.}
Once an IR is designed and implemented,
the static analysis implementation can essentially be regarded as IR manipulations.
It is promising to perform a ``postprocessing'' stage on such IR manipulations.  
Specifically,
this stage aims to understand the manipulation behavior, and attempt to optimize the IR manipulation in terms of performance.
Existing works have explored fusing several AST traversals into one traversal pass~\cite{Sakka2017recursive, Sakka2017recursive}.
Future works can consider more optimizing approaches as well as more kinds of IR manipulations.

\bibliographystyle{acm}
\bibliography{ref, voc}

\end{document}